\documentclass[twocolumn]{aa}

\usepackage[T1]{fontenc}
\usepackage{ae,aecompl}
\usepackage{lipsum}
\usepackage{lscape}
\usepackage{natbib}
\usepackage{listings}

\usepackage{realboxes}
\bibpunct{(}{)}{;}{a}{}{,} 

\usepackage{graphicx}	
\usepackage{amsmath}	
\usepackage{amssymb}	
\usepackage[dvipsnames]{xcolor}
\usepackage{txfonts}
\usepackage[colorlinks=true,allcolors=cyan]{hyperref}
\usepackage{ulem}
\usepackage{multirow}

\defcitealias{zhang2024hot}{Z+24}
\defcitealias{bekhti2016hi4pi}{HI4PI collaboration et al. 2016}
\defcitealias{ade2016planck}{Planck Collaboration XIII 2016}
\newcommand{\MSUN}{{\rm M}_\odot}

\begin{document} 
   \title{Quantifying Observational Projection Effects with a Simulation-based hot CGM model}
   
    \author{Soumya Shreeram\inst{1}\thanks{\href{shreeram@mpe.mpg.de}{shreeram@mpe.mpg.de}}, Johan Comparat\inst{1}, Andrea Merloni\inst{1}, Yi Zhang\inst{1}, Gabriele Ponti\inst{1,2}, Kirpal Nandra\inst{1}, John~ZuHone\inst{3}, Ilaria Marini\inst{4}, Stephan Vladutescu-Zopp\inst{5},  Paola Popesso\inst{3}, Ruediger Pakmor\inst{6}, Riccardo Seppi\inst{7}, Celine~Peroux\inst{3}, and Daniele Sorini\inst{8}}
    
    \institute{Max Planck Institute for Extraterrestrial Physics (MPE), Gie\ss enbachstraße 1, 85748 Garching, Munich, Germany 
    \and 
        INAF-Osservatorio Astronomico di Brera, Via E. Bianchi 46, I-23807 Merate (LC), Italy 
    \and
        Smithsonian Astrophysical Observatory, Observatory Building E, 60 Garden St, Cambridge, MA 02138, United States 
    \and
        European Southern Observatory, Karl-Schwarzschild-Stra\ss e 2, 85748 Garching, Munich, Germany
    \and
        Universit\" ats-Sternwarte M\"unchen, Scheinerstra\ss e 1, 81679 München, Germany
    \and 
        Max Planck Institute for Astrophysics,  Karl-Schwarzschild-Stra\ss e 1, 85748 Garching, Munich, Germany
    \and
        University of Geneva, 1205 Geneva, Switzerland
    \and
        Institute for Computational Cosmology, Department of Physics, Durham University, South Road, Durham, DH1 3LE, United Kingdom
             }
    \date{Received 17 September 2024; Accepted 27 March 2025; published 5 May 2025}

\abstract
  {}  
  {The hot phase of the circumgalactic medium (CGM) allows us to probe the inflow and outflow of gas responsible for dictating the evolution of a galaxy's structure. Studying the hot CGM sheds light on the physical properties of the  gas phase of the baryons, which is crucial to inform and constrain simulation models. With the recent advances in observational measurements probing the hot CGM in X-rays and thermal Sunyaev-Zeldovich (tSZ), we have a new avenue for widening our knowledge of gas physics and feedback.}
  {In this paper, we use the TNG300 hydrodynamical simulations to build a fully self-consistent forward model for the hot CGM. In order to do that, we construct a lightcone and generate mock X-ray observations of the large-scale structure. We quantify the main projection effects impacting CGM measurements, namely the locally correlated large-scale structure in X-rays and the effect due to satellite galaxies misclassified as centrals, which affect the measured hot CGM galactocentric profiles in stacking experiments.}
  {We present an analytical model that describes the intrinsic X-ray surface brightness profiles of halos across the stellar and halo mass bins. The increasing stellar mass bins result in decreasing values of $\beta$, the exponent quantifying the slope of the intrinsic galactocentric profiles. We  measure the effect of misclassified centrals in stacking experiments for three stellar mass bins  $10^{10.5-11} \MSUN$,  $10^{11-11.25} \MSUN$,  and $10^{11.25-11.5} \MSUN$. We find that the contaminating effect of the misclassified centrals on the stacked profiles increases when the stellar mass decreases. When stacking galaxies of Milky-Way-like stellar mass, this effect is dominant already at a low level of contamination: in particular, misclassified centrals contributing 30\%, 10\%, or 1\% of a sample dominate the measured surface brightness profile at radii $\geq 0.11 \times R_{500c}$, $\geq 0.24 \times R_{500c}, $ and $\geq 1.04 \times R_{500c}$, respectively.}
   {}
  
    \titlerunning{Quantifying Observational Projection Effects with a Simulation-based hot CGM model}
    \authorrunning{S. Shreeram}
   \maketitle

\section{Introduction}

The circumgalactic medium (CGM) plays a crucial role in interfacing the gas between the interstellar medium (ISM) within galaxies and the external intergalactic medium (IGM) outside them. The CGM is commonly defined as the gas within the virial radii of the galaxy and outside their stellar disks. It is the gravitationally bound gas that encompasses the fossil imprints of the physical mechanisms, such as outflows, inflows, and feedback processes, dictating the evolution of the galaxy~(see \citealt{tumlinson2017circumgalactic} for a review). These mechanisms are sensitive to the galaxy's environment and its halo properties, as shown by the complex Stellar-to-Halo-Mass-Relation (SHMR); see the review from \citet{Wechsler2018halo}. According to most models, the SHMR indicates that the low-halo-mass end is sensitive to stellar and supernovae (SN) driven feedback ~\citep{dekel2003feedback, Benson2003lf}, causing heating of the gas, hot bubbles \citep{mckee1977theory}, galactic wind outflows \citep{dekel1986origin}, and turbulence \citep{ostriker2011maximally, strickland2009supernova}. In contrast, Active Galactic Nuclei (AGN) should be the main drivers of feedback in the high-halo-mass end~\citep {Silk1998, 2012ARA&A..50..455F, eckert2021feedback}. Of particular interest is the peak of the SHMR relation at the pivotal halo mass $M_{\rm h}\sim 10^{12}$ M$_\odot$, similar to our Milky-Way (MW) mass, corresponding to the mass scale where star formation efficiency reaches its maximum. Thus, studying the volume-filling hot gas component of the CGM, especially in the MW halo mass regime, encapsulating a range of physical processes, is crucial for testing galaxy formation models \citep{faucher2023key}.

The most general scenario for the presence of such hot CGM assumes that infalling gas within halos $\gtrsim 10^{12} \MSUN$ is shock heated up to the virial temperature, $T_{\rm vir} \gtrsim 10^{6}$ K, resulting in X-ray emission~\citep{white1978core}. This hot gas is probed with the thermal Sunyaev-Zeldovich \citep[tSZ]{sunyaev1972formation} effect at mm-wavelengths~\citep{lim2021properties, Das:2023vs, Oren:2024aa} and in X-rays via absorption and emission. X-ray absorption studies use sightlines with bright background sources to probe the hot CGM via absorption lines like OVI or OVII K$\alpha$~\citep{galeazzi2007xmm, bhattacharyya2023hot, mathur2023probing}. The major challenge in this technique is constructing statistical samples and mapping the large-scale extent of the CGM, which will improve with future X-ray telescopes with microcalorimeters~\citep{Wijers2020eagle, bogdan2023circumgalactic}. On the emission side, studies use narrow-band and broad-band observations. The narrow-band emission comprising of OVII and OVIII metal lines dominates the lower halo mass regime $10^{12}-10^{13}$~M$_{\odot}$~\citep{Bertone2010oviii,van2013owl,Wijers2022lines} and is well-studied within our MW~\citep{Zheng2024mw, locatelli2024warm, ponti2023abundance, koutroumpa2007ovii}. However, for extragalactic studies, future high-spectral resolution instruments are required to distinguish the extragalactic emission from that of the MW foreground~\citep{Nelson:2023ui, Truong2023MNRASline, schellenberger2024mapping, zuhone2024properties}.

Multiple extragalactic single-object broad-band X-ray emission studies have been published~\citep{bogdan2013detection, bogdan2013hot, anderson2016deep, bogdan2017probing, li2017circum, das2019evidence}. However, due to the X-ray emissivity scaling with the square of the density, detections are limited to the densest parts of the CGM, within $40-150$ kpc from the galaxy centre. To map the extended CGM out to the halo virial radius, $R_{\rm vir}$,  \cite{anderson2015unifying} stacked SDSS galaxies with the full-sky X-ray data from the ROSAT survey. With the advent of SRG/eROSITA, we have unprecedented statistics for studying the hot CGM in X-rays with stacking analysis, as first shown by \citet{oppenheimer2020eagle}, who generated mock observations with IllustrisTNG and EAGLE hydrodynamical simulations. \citet{comparat2022erosita} and \citet{Chadayammuri:2022us}, and more recently, \citet[hereafter Z+24]{zhang2024hot} conduct such experiments with the eROSITA data and detect the hot CGM out to the virial radius of Milky-Way and M31-like galaxies. X-ray stacking analysis provides the most sensitive state-of-the-art observations of the hot CGM in the current observational landscape. Still, they are subject to various observational effects that affect the interpretation of the hot CGM. Consequently, we need a complete theoretical framework for describing the intrinsic hot CGM emission to disentangle the intrinsic signal from other observational effects.

A plethora of theoretical models (e.g., \citealt{Faerman:2017aa, voit2019circumgalactic, pal2019multiphase,  stern2024accretion, singh2021constraints, pandya2023unified, Faerman:2023aa}) have been used to describe and predict the intrinsic properties of the hot CGM; \cite{singh2024comparison} discuss the comparison of some idealised cases. Similarly, \citet{oppenheimer2020eagle, vladutescu2024radial} made explicit predictions of the intrinsic hot CGM profiles obtained by assigning mock X-ray emission to the hot halos found in cosmological hydrodynamical simulations. However, before testing theoretical and numerical predictions of the hot CGM against state-of-the-art observations, we must quantify all the effects that influence the hot CGM observations in stacking experiments. Among the primary sources of contamination in X-ray stacking experiments, inhibiting us from retrieving the physical properties of the detected CGM emission, are (1) the unresolved AGN and X-ray binaries~(XRB) population of galaxies~\citep{Biffi:2018wj, Vladutescu-Zopp:2023vo}, and (2) the projection effects within the hot gas emission of the cosmic web. These projection effects include the contribution from the Large-Scale Structure (LSS) halo environment (locally correlated X-ray emission), the effect of satellite galaxies being misclassified as central due to limitations in the (photometric) redshift accuracy for the galaxies in large surveys (see, e.g.,~\citealt{weng2024physical}), the offset between the X-ray centre and the centre defined by the minimum of the halo potential, and the Line-of sight (LoS) projection of uncorrelated X-ray emission of fore- and background structures. \citetalias{zhang2024hot} tried to model these effects empirically for the first time; however, we need forward models based on cosmological hydrodynamical simulations for a complete description and full understanding of these contaminating effects.

In this paper, we analyse numerical simulations to empirically quantify the effects of the locally correlated X-ray emission and the misclassified centrals, also called the satellite-boost effect, relevant for measuring the hot CGM in stacking experiments. To do this, we use the TNG300 hydrodynamical simulations \citep{pillepich2018first, marinacci2018first, naiman2018first, nelson2015illustris, springel2018first} to construct a lightcone and generate mock X-ray observations. We focus here on the projection effects of hot gas affecting the X-ray surface brightness profiles. The structure of the paper is as follows. Sec.~\ref{sec:lc} details the construction of the TNG300 lightcone (LC-TNG300) used for modelling the hot CGM in this study. We explain all the projection effects in detail in Sec.~\ref{subsec:projectioneffects}. In Sec.~\ref{sec:methods}, we describe the process for generating mock X-ray observations that mimic observational data and provide the machinery for computing the surface brightness profiles. Sec.~\ref{sec:results} presents the main results of this work done with the TNG300 lightcone. Lastly, Sec.~\ref{sec:disandcon} discusses the main findings and Sec.~\ref{sec:summary} provides an overview with prospects.

\begin{table*}[]
\centering
\caption{Stellar mass bins used to generate the X-ray surface brightness profiles.}
\label{tab:stellar_mass_bins}
\resizebox{0.75\textwidth}{!}{%
\begin{tabular}{cccccccc}
\hline \hline 
  \multicolumn{1}{l}{\begin{tabular}[c]{@{}c@{}}  $\log_{10}  
   \left(\frac{M_{\star}}{M_\odot} \right)$ \\ min-max  \end{tabular}} &

  \multicolumn{1}{l}{\begin{tabular}[c]{@{}c@{}}  $ \log_{10} \frac{M_{\star}}{M_\odot} $ \\ mean \end{tabular}} &
  
  \multicolumn{1}{l}{\begin{tabular}[c]{@{}l@{}}
  N$_{\rm centrals}$\\  \lstinline{CEN}$^{\rm sim}$ \end{tabular}} &
  
  \multicolumn{1}{l}{\begin{tabular}[c]{@{}l@{}}N$_{\rm satellites}$\\ \lstinline{SAT}$^{\rm sim}$   \end{tabular}} &
  
  \multicolumn{1}{l}{\begin{tabular}[c]{@{}c@{}}$  \log_{10} \frac{M_{200m}}{M_\odot}  $  \\ mean \end{tabular}} &
  
  \multicolumn{1}{l}{\begin{tabular}[c]{@{}c@{}}  $\log_{10} \left( \frac{M_{200m}}{M_\odot}  \right) $ \\ min-max  \end{tabular}}&
  
  \begin{tabular}[c]{@{}l@{}} $ {R_{\rm 200m}}  $ \\ kpc \end{tabular} &
   \begin{tabular}[c]{@{}l@{}} $ {R_{\rm 500c}}  $ \\ kpc \end{tabular} 
  \\ \hline 
  
$10.5-11$    & $10.7$ & $5109$         & \multicolumn{1}{l}{$2719$}     & $12.7$ &  $[11.9, 13.7]$   & $525.02$  & $242.75 $\\
$11-11.25$   & $11.1$ & $680$          & \multicolumn{1}{l}{$220$}       & $13.3$ &  $[12.6, 14.0]$   & $816.05$ & $369.01$\\
$11.25-11.5$   & $11.4$ & $305$          & \multicolumn{1}{l}{$60$}       & $13.6$ &  $[13.2, 14.2]$   & $1077.71$ & $484.52$\\
\end{tabular}%
}
\tablefoot{We present the mean stellar mass within twice the stellar half mass radius, the total number of central galaxies, the total number of satellite galaxies, mean halo mass, the minimum and maximum values of halo masses for the given stellar mass bin, the mean $R_{200m}$, and the mean $R_{500c}$.}
\end{table*}

\begin{table*}[]

    \centering
    \caption{Halo mass bins used to generate the X-ray surface brightness profiles.}
    \label{tab:halo_mass_bins}
\resizebox{0.6\textwidth}{!}{%
    \begin{tabular}{lcccccc}
    \hline \hline 
      \multicolumn{1}{l}{\begin{tabular}[l]{@{}l@{}}  $\log_{10} \rm \frac{M_{200m}}{M_\odot}$  \\ min-max  \end{tabular}} &
      $\log_{10}  \rm \frac{M_{200m}}{M_\odot} $ &
      \multicolumn{1}{l}{\begin{tabular}[c]{@{}l@{}}N$_{\rm Distinct\ halos}$\\  \lstinline{CEN}$^{\rm sim}_{\rm halo}$ \end{tabular}} &
      $\log_{10}  \rm \frac{M_{\star, <2\times R_\star}}{M_\odot} $ &
      {\begin{tabular}[c]{@{}l@{}} $ {R_{\rm 200m}}  $ \\ kpc \end{tabular}} &
      {\begin{tabular}[c]{@{}l@{}} $ {R_{\rm 500c}}  $ \\ kpc \end{tabular}}  \\ \hline 
    $12.5-13.0$  & $12.7$ & \multicolumn{1}{l}{$3, 677$} & $10.7$ & $375.54$ & $246.35 $\\
    $13.0-13.5$  & $13.2$ & \multicolumn{1}{l}{$1, 113$} & $11.0$ & $549.27$& $353.35$\\
    $13.5-14.0$  & $13.7$ & \multicolumn{1}{l}{$382$}   & $11.4$ & $810.09$ & $519.34$\\
    \end{tabular}%
}
\tablefoot{We present the mean halo mass, the total number of distinct halos, the mean stellar mass, the mean $R_{\rm 200m}$ and the mean $R_{\rm 500c}$.}
\end{table*}

\section{Simulated lightcone with IllustrisTNG: LC-TNG300}
\label{sec:lc}

\begin{figure*}[h]

    \centering
    \includegraphics[width=\linewidth]{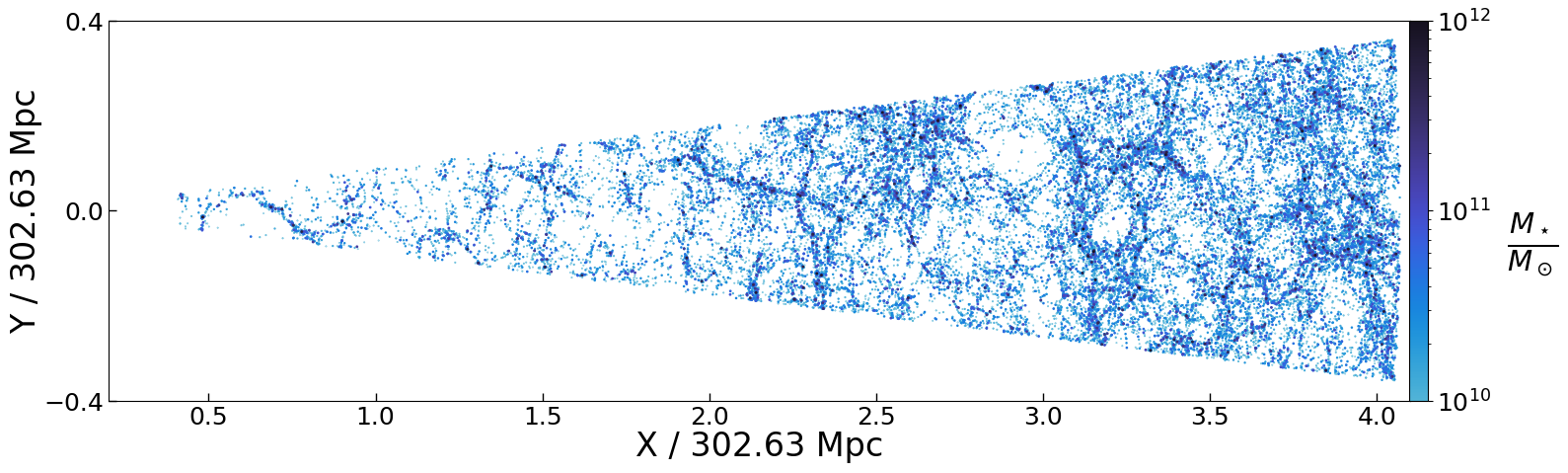}
    \caption{Illustration of the lightcone built using TNG300 in the x-y plane. The figure shows the subhalos within LC-TNG300 at $0.03\leq z\leq0.3$ remapped using \texttt{boxremap}~\citep{carlson2010embedding}. The observer is set at the $(0, 0, 0)$ location. The lightcone goes out to $1231\rm \ cMpc$ along the x-axis, subtending an area of $47.28\rm \ deg^2$ on the sky in the y-z plane. The subhalos are colour-coded with their stellar masses.}
     \label{fig:lightcone_image}
\end{figure*}

\begin{figure*}

    \includegraphics[width=\linewidth]{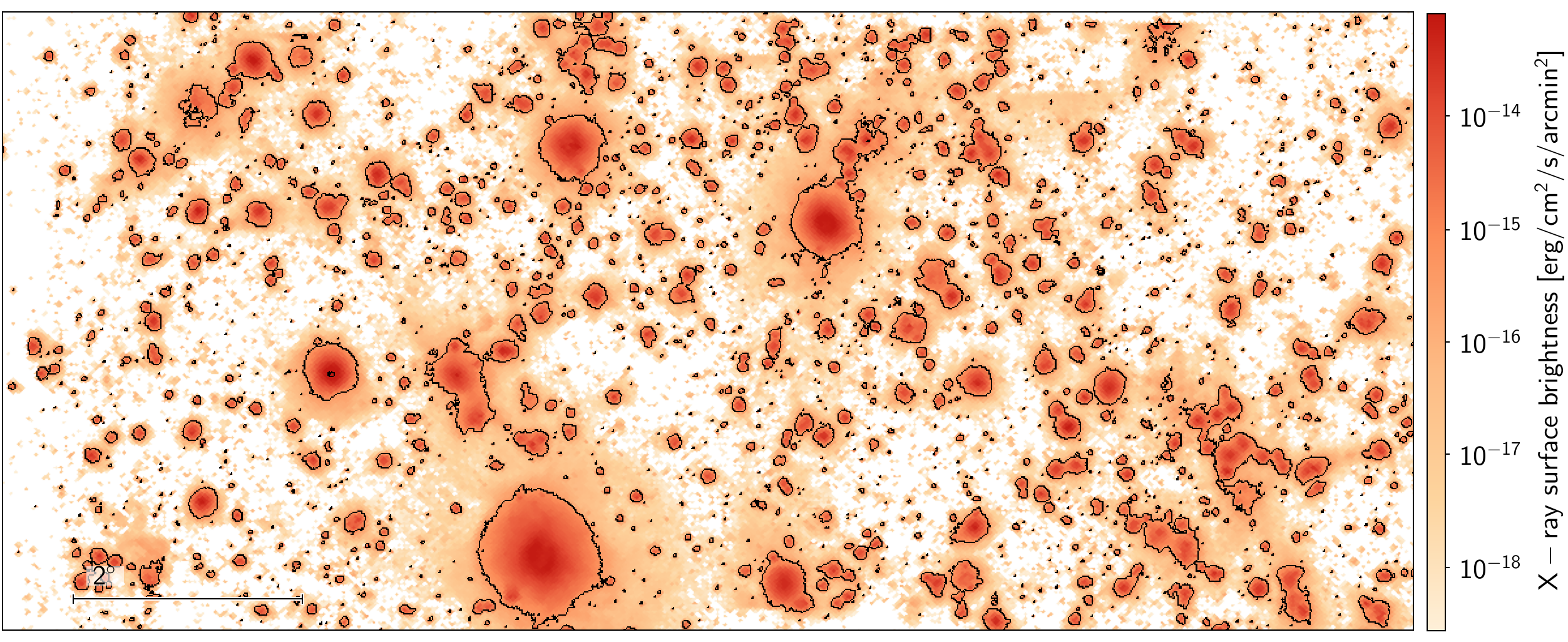}
    
    \caption{Projected rest-frame X-ray events from the TNG300 ligthcone in the $0.5-2.0$ keV band for a telescope with energy-independent collecting area $1000$ cm$^2$ and exposure time of $1000$ ks. The events are generated using the hot gas cells within the TNG300 lightcone at redshifts $0.03 - 0.3$ using \texttt{pyXsim}~\citep{zuhone2016pyxsim}. The centre for the projection onto the sky is chosen as R.A., Dec.$\equiv (0., 0.)$ degrees. The contours represent the two X-ray surface brightness levels of $5\times 10^{-14}$ erg s$^{-1}$ cm$^{-2}$ arcmin$^{-2}$ and $2\times 10^{-16}$  erg s$^{-1}$ cm$^{-2}$ arcmin$^{-2}$. \label{fig:allevents}}
\end{figure*}

\begin{figure*}[t]

    \centering
    \includegraphics[width=\linewidth]{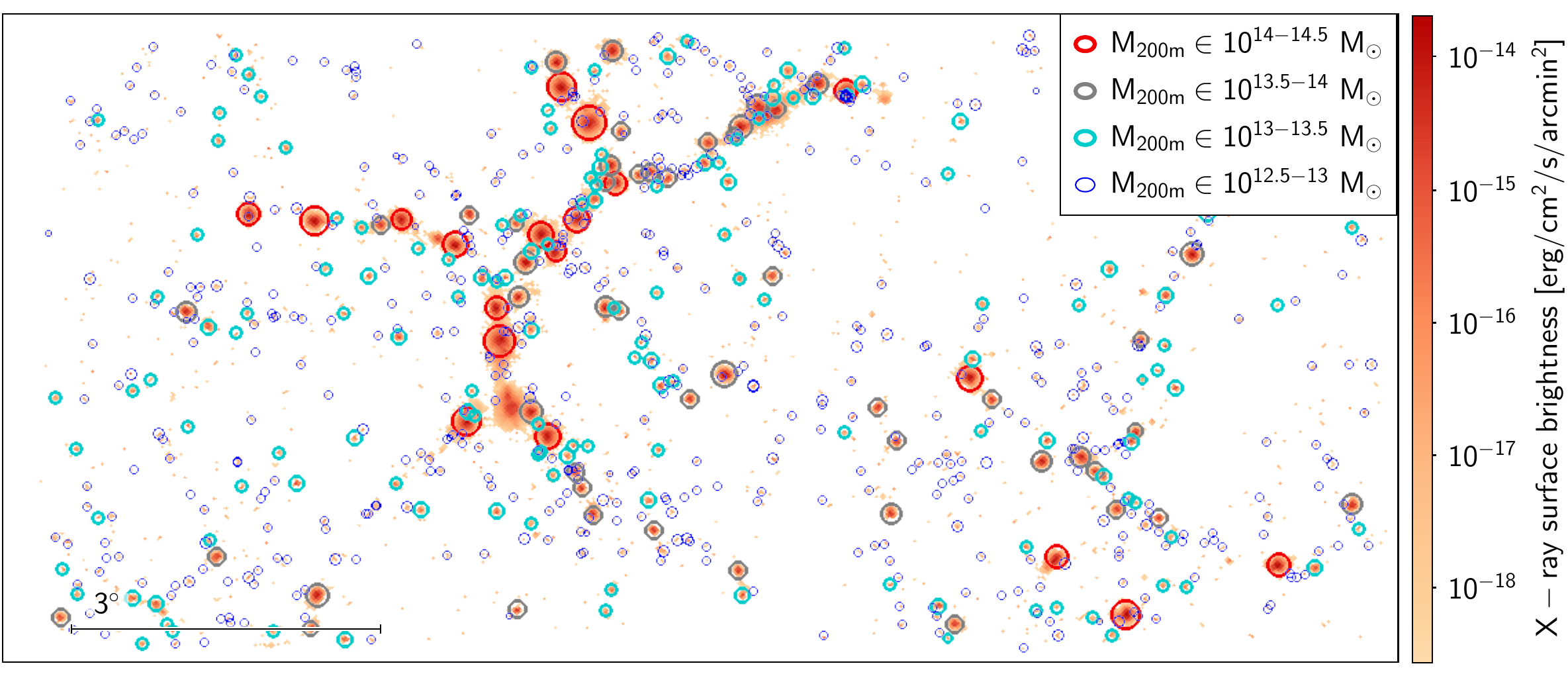}
    
    \caption{Projected halos ({centrals}) are overplotted with their corresponding scaled $R_{500c}$ at $z=0.3$; the $R_{500c}$ of the halos are represented by the size of the circles. We also show the projected rest-frame X-ray events from the TNG300 ligthcone in the $0.5-2.0$ keV band for a telescope with energy-independent collecting area $1000$ cm$^2$ and exposure time of $1000$ ks at the redshift slice of $0.284\leq z\leq0.3$. The halos with $M_{200m} \in 10^{14-14.5}\ \rm M_\odot$ are shown in red, $M_{200m} \in 10^{13.5-14}\ \rm M_\odot$ in grey, $M_{200m} \in 10^{13-13.5}\ \rm M_\odot$ in cyan, and  $M_{200m} \in 10^{12.5-13}\ \rm M_\odot$ in blue. \label{fig:allevents2}}
\end{figure*}

Since we are interested in studying the impact of the locally correlated X-ray emission within the LSS environment of the halo, we require a simulation that contains all the complexities introduced by feedback and cooling, which are imprinted in X-ray measurements. We also require a large box size to encompass the effects of the cosmological LSS itself. Therefore, we use cosmological hydrodynamical simulations that self-consistently predict the LSS and its impact on gas dynamics within a halo. We use the IllustrisTNG cosmological hydrodynamical simulation with the box of side length $302.6$ Mpc~\citep[TNG300;][]{nelson2019illustristng}\footnote{\url{ http://www.tng-project.org} }. IllustrisTNG accounts for many physical processes, among which the most notable ones are star formation regulated by a subgrid ISM model~\citep{springel2003cosmological}, metal enrichment~\citep{naiman2018first}, radiative gas cooling, galactic wind outflows~\citep{pillepich2018simulating}, magnetic fields and diffuse radio emissions~\citep{marinacci2018first}, supermassive black hole growth with Bondi accretion and mergers, thermal and kinetic modes for black hole feedback~\citep{weinberger2018supermassive}. IllustrisTNG also reproduces correlation functions and power spectra of particles and halos~\citep{springel2018first}. TNG300 contains $2500^3$ dark matter particles, with a baryonic mass resolution of $1.1\times 10^7 \rm \ M_{\odot}$, a comoving value of the adaptive gas gravitational softening length for gas cells of $370$ comoving parsec, gravitational softening of the collisionless component of $1.48$ kpc, and dark matter mass resolution of $5.9\times 10^7\ \rm M_{\odot}$. The TNG simulations adopt the \citetalias{ade2016planck} cosmological parameters, with the matter density parameter $\rm \Omega_m = \Omega_{dm} + \Omega_{b} = 0.3089$, baryonic density parameter $\Omega_b = 0.0486$, Hubble constant ${\rm H}_{0} = 100 h $ km/s/Mpc with $h = 0.6774$, and $\Omega_\Lambda = 0.6911$.

The Friends-of-Friends (FoF) algorithm is applied to the dark matter particles with linking length $b=0.2$ to obtain the halos. The subhalos are retrieved with \textsc{subfind}~\citep{springel2001populating, dolag2009substructures}, which detects gravitationally bound substructures that are equivalent to galaxies in observations. Additionally, \textsc{subfind} also classified the subhalos into centrals and satellites, where centrals are the most massive substructure within a distinct halo. 

To model the hot gas emission in the redshift range $0.03 \lesssim z \lesssim 0.3$, as done in observations (e.g. \citealt{comparat2022erosita, Chadayammuri:2022us, zhang2024hot}), we need four boxes lined up behind each other to arrive at the comoving distance, $\rm d_C(z=0.3)$, of $1231\rm \ Mpc$. For a full-sky lightcone going up to $z=0.3$, we must replicate and stack 512 boxes; however, in this technique, the large-scale projection effects cannot be estimated properly; see~\cite {merson2013lightcone}. Therefore, we box remap\footnote{\url{http://mwhite.berkeley.edu/BoxRemap/}} all snapshots into a configuration where the longest length of one of the sides is $\sim4\times$ the original length~\citep{carlson2010embedding}. This technique ensures that the new elongated box has a one-to-one remapping, remains volume-preserving, and keeps local structures intact. For a box whose original dimensions are normalized to $(1, 1, 1)$, the unique solution for the transformed box sides is $(4.1231, 0.7276, 0.3333)$, respectively. We remap the coordinates of particles (gas, dark matter, and stars) and the halo and their subhalo catalogues. There are 22 snapshots within the observationally motivated redshift range of $0.03 \leq z \leq 0.3$. Therefore, we obtain $22$ remapped particle cuboids and $22$ remapped galaxy catalogue cuboids at redshift $z<0.3$. We define the observer's position as being at a corner of the smallest face. The opening angles for the observer are $(\theta_{\rm LC\ obs}, \phi_{\rm LC\ obs}) = (10.16,\ 4.64)$ degrees.  The area subtended on the sky for a given observer is $47.28\rm \ deg^2$. 
We illustrate the lightcone constructed with the box remap technique in Fig.~\ref{fig:lightcone_image}. We also publically release \texttt{LightGen}, the code used for generating the lightcone in this work.

The distinct halos, \lstinline{CEN}$^{\rm sim}_{\rm halo}$, and subhalos within the LC-TNG300 are catalogued with their physical properties and binned in stellar mass and halo mass bins, as shown in Tab.~\ref{tab:stellar_mass_bins} and Tab.~\ref{tab:halo_mass_bins}. We have $5,109$ centrals and $2,719$ satellites of stellar mass in the Milky Way range ($\rm M_\star = 10^{10.5-11} \MSUN$), which is sufficient to statistically model the projection effects in these halos. 

This paper defines the stellar mass used from TNG300 as the mass within twice the stellar half-mass radius. We present quantities relative to both critical and mean densities because $R_{500c}$\footnote{$R_{\rm 500c}$ is the radius at which the density of the halo is  $500\times$ the critical density of the universe.} is the radius most commonly used by X-ray astronomers (see e.g., \cite{lyskova2023x} and references therein), and $R_{200m}$\footnote{$R_{\rm 200m}$ is the radius at which the density of the halo is  $200\times$ the mean matter density (cold dark matter and baryons).} represents the halo's viral radius and is theoretically more relevant as it presents quantities within the virialized halo. The galaxy catalogue is divided into central, \lstinline{CEN}$^{\rm sim}$, and satellites, \lstinline{SAT}$^{\rm sim}$ for the concerned stellar mass bins. To construct the \lstinline{SAT}$^{\rm sim}$ catalogue, we refer to the halo/subhalo classification, where we match all the central galaxies with the distinct halos, thereby leaving behind all the secondary subhalos or satellites.

\section{Contributions to observed X-ray surface brightness profiles}\label{subsec:projectioneffects}

The main objective of this study is to model the various sources of contamination that come into play when measuring the hot gas component of the CGM. Here, we list all the components contributing to the observed X-ray surface brightness profile.
\begin{enumerate}
    \item {Intrinsic emission from the halo}. This corresponds to the emission within the radius, $R_{\rm 200m}$, of the central halo in 3D. We associate this X-ray emission with being intrinsic to the galaxy CGM. It must be noted that the definition of a halo boundary at which the gas is bound is non-trivial; we refer to~\cite{diemer2017splashback2} and references therein for more details. 

    \item {Locally correlated environment}. This component encapsulates the surrounding emission of the galaxy with the LSS in which it resides. Also known as the 2-halo term~\citep{Cooray2002halo, kravtsov2004dark}, this corresponds to the contribution arising due to the local background changing with the size of the halo. This effect is explored in tSZE studies, where \cite{vikram2017measurement}, \cite{lim2021properties} show the impact of the two-halo component, where the one-halo contribution at mass scales $M_{\rm 200} \leq 10^{13-13.5}$ h$^{-1}$ M$_\odot$ is swamped by the two-halo term due to nearby massive systems dominating the measured signal. Given our focus on MW-size halos, modelling the contribution from the locally-correlated background is crucial for disentangling intrinsic hot CGM emission from the local background.
    
    \item {Intrinsic emission from satellites}. The X-ray emission from the satellite galaxy contributes to the total observed CGM emission in projection. As the host galaxy is more massive, with a deeper potential well, the contribution of this effect is negligible in stacking experiments (see \cite{rohr2024sat} for further insights on the detectability of the satellite emission).
    
    \item {Contamination from misclassified centrals}. In observations that use photometric surveys, classifications between centrals and satellites are inhibited due to limitations in the redshift accuracy. This effect is large and unavoidable for photometric surveys ($\sim 30\%$ contaminants for MW-like galaxies; Sec 3.5 in~\citetalias{zhang2024hot}) and significantly mitigated for spectroscopic surveys ($\sim 1\%$ contaminants for MW-like galaxies; Sec 3.7 in~\citetalias{zhang2024hot}), but not completely removed due to survey incompleteness or residual uncertainty in the central/satellite classification for systems with a low number of galaxies (see also ~\citealt{weng2024physical} who quantify the effect for absorbers in cold gas with TNG50). This implies that the measured X-ray surface brightness profiles in stacked samples of galaxies classified as centrals contain the intrinsic emission around truly central galaxies but are contaminated by the emission measured around satellites (off-centred) misclassified as centrals. In conclusion, including the emission around satellites in the stacking analysis alters the recovered profiles and, therefore, must be modelled.

    \item {The X-ray to true centre offset}. The theoretically defined centre of a halo, which is the minimum of the dark matter potential well, could be offset from the observationally defined centres, e.g., the peak of the SZ signal or the X-ray emission, or the optical/infrared centre inferred from the stellar distribution ~\citep{zitrin2012miscentring, lauer2014brightest, saro2015constraints, cui2016does, Seppi2023offset}. This offset is associated with physical effects such as galaxy mergers~\citep{martel2014major}, misidentification of the brightest halo galaxy~\citep{Hoshino2015lrgs,hikage2018testing,oguri2018optically, zhang2019dark} or failure of the brightest halo galaxy being a proxy for the minimum of the halo potential in optical studies~\citep{skibba2011brightest}. Overall, miscentering causes a smoothing effect in stacking studies~\citep{Oguri2011smooth}, and therefore, must be calibrated and modelled. This effect of miscentering correlates with halo mass; hence, it is well-studied at the cluster and group scales. As we simulate X-ray emission around lower mass, MW-like halos in this work, we extend the discussion of miscentering of the X-ray emission from the minimum of the host halo potential to MW mass scales (see discussion in Sec.~\ref{subsec:xray2true_cen}).

    \item {Contamination from other X-ray sources}. Other X-ray emitting sources like XRB and AGN contaminate the measured hot CGM. XRB emission is distributed on the scale of the stellar body of a galaxy; however, for an instrument with a point spread function (PSF) like eROSITA, it is unresolved and appears as a point source. \citetalias{zhang2024hot} take special precautions to account for this by masking the eROSITA detected point sources within the X-ray data. They also model additional contributions from the unresolved XRB and AGN. For this study, we leave the modelling of the AGN and XRB resolved/unresolved emissions with the LC-TNG300 for future work.

    \item {Line of sight (LoS) projection}. Objects along the line of sight, which do not reside near the galaxy, also contaminate the detected signal. Nevertheless, this uncorrelated contamination is well-modelled as the large-scale background and foreground in observations and, therefore, is not discussed further in this work. 

    \item {Other X-ray background components}. The local flat background is a combination of (1) the Soft X-ray background (SXRB; \citealt{henley2010xmm, henley2012xmm, henley2013xmm, nakashima2019x, gupta2023mw, Pan2024swcx, yeung2024srg}), which is composed of emission from the local hot bubble, the MW CGM, other intervening galactic structure and the time variable solar-wind charge exchange, (2) the Cosmic X-ray background~\citep{de20042, gilli2007synthesis, luo2016chandra, cappelluti2017probing}, which is dominated by AGN, and (3) the instrumental effects. In this study, we simulate intrinsic X-ray events from hot gas alone; therefore, this study does not expand on the impacts of the other well-studied X-ray background components.
\end{enumerate}

This work focuses on modelling the locally correlated environment and the effect due to misclassified-centrals (the satellite-boost effect). We also quantify the effect of the X-ray to true centre offset for the halo mass bins considered in this work.

\section{Method}
\label{sec:methods}

We detail the process to create mock X-ray observations in Sec.~\ref{subsec:mockX} and introduce the formalism used to fit the X-ray surface brightness profiles in Sec.~\ref{subsec:profiles}. We detail the data products generated to study the projection effects in Sec.~\ref{subsec:dataproducts}.

\begin{figure*}

    \centering
    \begin{minipage}{\textwidth}
        \includegraphics[width=\linewidth]{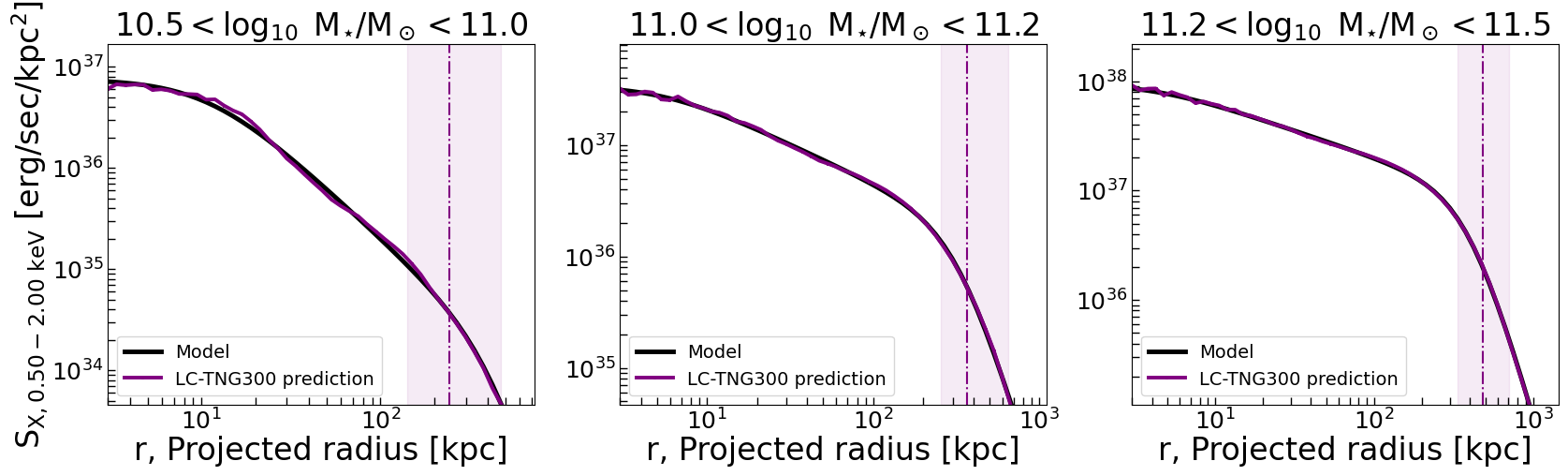}
    \end{minipage}%
    \caption{Mean X-ray surface brightness profiles in the stellar mass bins: $M_{\star} = 10^{10.5-11}\ \MSUN$, corresponding to MW-like galaxies ({left}), $M_{\star} = 10^{11-11.25}\ \MSUN$ ({centre}), and $M_{\star} = 10^{11.25-11.5}\ \MSUN$ ({right}). The vertical dashed line is the mean $R_{500c}$ at $242.75$ kpc ({left}), $369.01$ kpc ({centre}), and $484.52$ kpc ({right}) of the respective galaxy stellar mass bin with the shaded area corresponding to the minimum and maximum values. The black line is the analytical model, shown in Eq.~\ref{eqn:modified_beta}, fit to the LC-TNG300 mean X-ray surface brightness profiles. The best-fitting parameters for the model are given in Tab.~\ref{tab:best-fit-params0}. We find decreasing values for the slope, $\beta$, with increasing stellar mass bins. \label{fig:best-fit}}
\end{figure*}

\begin{figure*}

    \centering
    \begin{minipage}{\textwidth}
        \includegraphics[width=\linewidth]{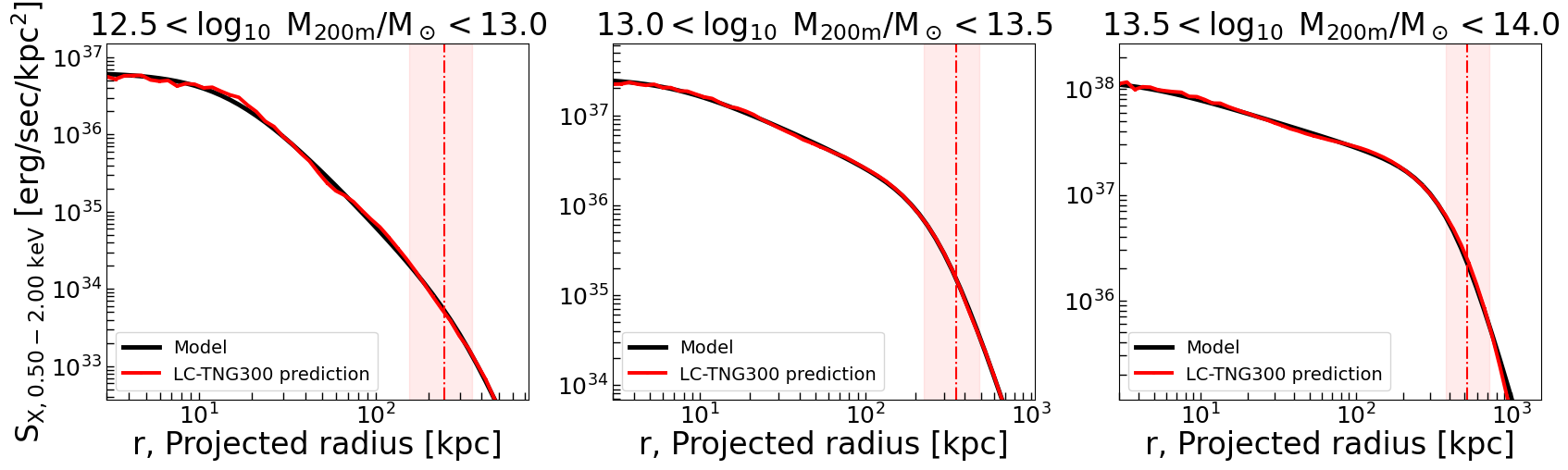}
    \end{minipage}%
    \caption{Mean X-ray surface brightness profiles in the halo mass bins: $M_{\rm 200m} = 10^{12.5-13}\ \MSUN$, corresponding to MW-like galaxies ({left}), $M_{\rm 200m} = 10^{13-13.5}\ \MSUN$ ({centre}), and $M_{\rm 200m} = 10^{13.5-14}\ \MSUN$ ({right}). The vertical dashed line is the mean $R_{500c}$ at $246.35$ kpc ({left}), $353.35$ kpc ({centre}), and $519.34$ kpc ({right}) of the respective halo mass bin with the shaded area corresponding to the minimum and maximum values. The black line is the analytical model, shown in Eq.~\ref{eqn:modified_beta}, that fits the LC-TNG300 mean X-ray surface brightness profiles. The best-fitting parameters for the model are given in Tab.~\ref{tab:best-fit-params1}.  We find decreasing values for the slope, $\beta$, with increasing halo mass bins. \label{fig:best-fit-halo}}
\end{figure*}

\begin{table*}[]

\centering
\caption{For every stellar mass bin, we present the best-fitting parameters of the model (see Eq.~\ref{eqn:modified_beta}).}
\label{tab:best-fit-params0}
\begin{tabular}{llllll}
\hline \hline
\multicolumn{1}{l}{\begin{tabular}[c]{@{}c@{}}$\log_{10} \rm{M}_\star/\MSUN$\\ min max\end{tabular}} &
$\log_{10} S_0$ &
\begin{tabular}[c]{@{}l@{}}$r_c$\\ {[}kpc{]}\end{tabular} &
$ \beta$ &
\begin{tabular}[c]{@{}l@{}}$r_s$\\ {[}kpc{]}\end{tabular} &
$\epsilon$ \\ \hline
10.5 - 11 & $ 36.963_{-0.001}^{+0.001}$ & $ 11.37_{-0.05}^{+0.05}$ & $ 0.443_{-0.001}^{+0.000}$ & $ 441_{-6}^{+4}$ & $ 4.5_{-0.1}^{+0.1}$ \\
11 - 11.25 & $ 37.631_{-0.002}^{+0.002}$ & $ 5.39_{-0.07}^{+0.06}$ & $ 0.283_{-0.001}^{+0.001}$ & $ 328_{-2}^{+2}$ & $ 4.24_{-0.05}^{+0.05}$ \\
11.25 - 11.5 & $ 38.080_{-0.003}^{+0.003}$ & $ 4.00_{-0.07}^{+0.08}$ & $ 0.249_{-0.001}^{+0.001}$ & $ 407_{-2}^{+2}$ & $ 4.70_{-0.03}^{+0.03}$
\end{tabular}
\vspace{2mm}
\centering
\caption{For every halo mass bin, we present the best-fitting parameters of the model (see Eq.~\ref{eqn:modified_beta}).}
\label{tab:best-fit-params1}

\begin{tabular}{llllll}
\hline \hline
\multicolumn{1}{c}{\begin{tabular}[c]{@{}c@{}}$\log_{10} \rm{M}_{\rm 200m}/\MSUN$\\ min max\end{tabular}} & $\log_{10} S_0$ & \begin{tabular}[c]{@{}l@{}}$r_c$\\ {[}kpc{]}\end{tabular} & $ \beta$ & $r_s$ & $\epsilon$ \\ \hline
12.5 - 13 & $ 36.804_{-0.001}^{+0.001}$ & $ 15.4_{-0.2}^{+0.2}$ & $ 0.575_{-0.001}^{+0.001}$ & $ 322_{-31}^{+33}$ & $ 3.0_{-0.4}^{+0.6}$ \\
13 - 13.5 & $ 37.414_{-0.002}^{+0.002}$ & $ 7.05_{-0.07}^{+0.08}$ & $ 0.307_{-0.001}^{+0.001}$ & $ 276.0_{-0.9}^{+0.9}$ & $ 4.80_{-0.03}^{+0.02}$ \\
13.5 - 14 & $ 38.085_{-0.002}^{+0.002}$ & $ 4.08_{-0.06}^{+0.07}$ & $ 0.242_{-0.001}^{+0.001}$ & $ 412.7_{-0.5}^{+0.5}$ & $ 4.894_{-0.002}^{+0.002}$
\end{tabular}
\tablefoot{: $S_0$, the central surface brightness; $r_{\rm c}$, the core radius; $\beta$, the exponent quantifying the slope of the profile; $r_{s}$, the scale radius at which the slope changes to $\epsilon$.}

\end{table*}
\subsection{Mock X-ray observation}
\label{subsec:mockX}

The photons are simulated in the $0.5-2.0$ keV intrinsic band with \texttt{pyXsim}~\citep{zuhone2016pyxsim}, which is based on \textsc{phox}~\citep{Biffi:2013uh}, by assuming an input emission model where the hot X-ray emitting gas is in collisional ionisation equilibrium. The spectral model computations of hot plasma use the Astrophysical Plasma Emission Code, \textsc{apec}\footnote{APEC link \url{https://heasarc.gsfc.nasa.gov/xanadu/xspec/manual/node134.html}} code~\citep{smith2001collisional} with atomic data from \textsc{atomdb} v3.0.9~\citep{foster2012updated}. This model requires the plasma temperature of the gas cells (in keV), the metal abundances, the redshift $z$ and the normalisation,
\begin{equation}
    N = \frac{10^{-14}}{4 \pi [D_{\rm A}(1+z)]^2} \int n_{\rm e} n_{\rm H} \rm{d}V,
\end{equation}
where $D_{\rm A}$ is the angular diameter distance to the source (cm), $ \rm{d}V$ is the volume element (cm$^3$), $n_{\rm e}$ and $n_{\rm H} $ are the electron and hydrogen densities (cm$^{-3}$), respectively. 
The temperature is calculated from the internal energy $u$ and the electron abundance $x_{\rm e} (= n_{\rm e}/n_{\rm H})$ of the gas cells\footnote{See FAQ here: \url{https://www.tng-project.org/data/docs/faq/}}. The temperature $T$ for every gas cell is defined as
\begin{equation}\label{eqn:temp}
    T = (\gamma -1)  \frac{u}{k_{\rm B}} \mu
\end{equation}
where $k_{\rm B}$ is Boltzmann's constant in CGS units and $\gamma=5/3$ is the adiabatic index. The mean molecular weight $\mu$ is given as
\begin{equation}
    \mu = \frac{4}{1+3 X_{\rm H} + 4 X_{\rm H}x_{\rm e}} m_{\rm p}
\end{equation}
where $X_{\rm H }=0.76$ is the hydrogen mass fraction and $m_{\rm p}$ is the proton mass in grams. The metal abundances within TNG are provided for the snapshots at redshift intervals of every $0.1$. As a result, 19 of the 22 snapshots within the lightcone constructed in this work lack metallicity information. Given the lack of metallicity information and the inaccuracies introduced by extrapolation of metallicity values between the 0.1 redshift intervals due to the evolution of gas particles, we assume a constant metallicity of $0.3\ Z_\odot$; this is consistent with measurements for our MW~\citep{miller2015constraining, bregman2018extended, kaaret2020disk, Ponti:2023aa}. This work uses the solar abundance values from~\cite{anders1989abundances}.

The TNG star formation model is based on the subgrid two-phase model proposed by \cite{springel2003cosmological}, with some modifications \citep[see][and references therein]{pillepich2018first}. The gas cells that emit efficiently in the soft X-ray band (i.e. [0.3–5] keV) are due to the SN-driven kinetic decoupled winds, which ultimately deposit energy into non-star-forming gas cells. Additionally, the hot component of this two-phase model exhibits typically high temperature ($> 10^5$ K). As further explained in \cite{truong2020x} (see Appendix B-C), because this multiphase structure within the TNG model is not resolved, and it is instead modelled by a simplistic two-phase structure with unrealistic assumptions, we cannot make a sensible estimate of the X-ray emission from the unresolved phases of the ISM. Therefore, by excluding the parameter space of the warm-neutral ISM~\citep{le2014towards, rahmati2016cosmic, wijers2019abundance}, namely: (1) excluding star-forming gas cells, (2) ignoring gas cells below $10^5$ K, and (3) ignoring gas cells with densities above $10^{-25}$ g/cm$^3$, we ensure the gas particles used in this work are physically emitting X-rays. 

Within \texttt{pyXSim}, the number of photons generated depends on the specified collecting area of the assumed X-ray telescope, its exposure time, and redshift. We generate sufficient photons by assuming a telescope with an energy-independent collecting area of $1000$ cm$^2$ (about 3/4 of the eROSITA field-of-view-average effective area at 1 keV) and an exposure time of $1000$ ks. The photon-list is generated in the observed frame of the X-ray emitting gas cells and is corrected to rest frame energies. The LoS direction determines the event's position in the sky. We define the LoS along the x-axis within the lightcone. By applying a $\theta_{\rm LC\ obs}/2=5.08$ deg rotation along the z-axis and $-\phi_{\rm LC\ obs}/2=-2.32$  deg rotation along the y-axis, we centre the y-z plane at $(0,0)$ degrees. The photons generated by the gas particles are projected onto the sky; the resulting image is shown in Fig.~\ref{fig:allevents}.

We also project the halo and subhalo positions on the sky, as shown in Fig.~\ref{fig:allevents2} for a redshift slice of $z=0.3$. The X-ray events are then stacked around the projected halo/subhalo positions. The surface brightness profiles\footnote{We define the surface brightness profiles in this work in units of erg/s/kpc$^2$, consistent with \citetalias{zhang2024hot}, also called the luminosity profile.} are calculated $\mathcal{S}(r)$ as a function of the projected radius on the sky within the $0.5-2$ keV energy band. The centre is chosen as the location of the most bound particle within the halo/subhalo as found by the FoF/\textsc{subfind} algorithm. The number of photons with rest frame energy $E$ [erg] in each radially outward bin $r$, $N(E, r)$, is weighted by the area, $\mathcal{A}$ [kpc$^2$], of the 2D shell between $r$ and $r+\rm{d} r$, the fraction of photons collected from the source, $f_{\rm A}$, and the exposure time, $t_{\rm exp}$ [seconds],
\begin{equation}\label{eq:sx_profile}
    \mathcal{S}_{\rm X}(r) = \frac{\sum_{E_i= 0.5\ \rm keV}^{2\ \rm keV}N(E_i, r)\ E_i}{f_{\rm A} t_{\rm exp} \mathcal{A}(r)}\ \Bigg[\frac{\rm erg}{\rm sec\ kpc^2}\Bigg].
\end{equation}
Here, $f_{\rm A}$ is the fraction of the source photons collected by the synthesized telescope that has a collecting area $A$:
\begin{equation}
    f_{\rm A} = \frac{A}{4\pi d_{\rm L}(z)^2}
\end{equation}
where $d_{\rm L}(z)$ is the luminosity distance to the source.

\subsection{X-ray surface brightness profiles and their analytic modelling}
\label{subsec:profiles}

The hot CGM surface brightness profile can be analytically described by the $\beta$ model~\citep{cavaliere1976x};
\begin{equation}\label{eqn:beta}
     \bar{\mathcal{S}}_{\rm X, 0.5-2.0\ keV}[r]   = S_0 \left[  1 + \left(\frac{r}{r_{\rm c}} \right)^2 \right]^{-3\beta + \frac{1}{2}},
\end{equation}
where $S_0$ is the central surface brightness, $r_{\rm c}$ is the core radius at which the profile slope becomes steeper, and $\beta$ is the exponent quantifying the slope of the profile. In cases where the outskirt steepens more than the slope defined for the inner radii by the $\beta-$model, following~\citet{vikhlinin2006chandra}, we introduce a new slope-parameter,
\begin{equation}\label{eqn:modified_beta}
     \bar{\mathcal{S}}_{\rm X, 0.5-2.0\ keV}[r]   = S_0  \left[  1 + \left(\frac{r}{r_{\rm c}} \right)^2 \right]^{-3\beta + \frac{1}{2}} \times  \left[  1 + \left(\frac{r}{r_{\rm s}} \right)^\gamma \right]^{-\epsilon/\gamma},
\end{equation}
where $r_s$ is the scale radius at which the slope changes to $\epsilon$ and $\gamma$ defines the width of the transition region. We fix $\gamma=3$ and restrict the priors on $r_{s}>r_{c}$ and $\epsilon<5$, as suggested by \cite{vikhlinin2006chandra}, in the fitting procedure.

\subsection{Prerequisite data products for quantifying projection effects}
\label{subsec:dataproducts}

To obtain contributions from the large-scale structure, the locally correlated X-ray emission, we generate cubes and profiles with the \lstinline{CEN}$^{\rm sim}$ catalogue for the two cases as follows:
    \begin{enumerate}
        \item {\lstinline{R}$_\mathbf{\rm 200m}$ cubes and profiles}:
        every central galaxy within the halo catalogue is assigned X-ray events within $R_{\rm 200m}$ of the parent halo. We define these profiles as the intrinsic hot gas emission profiles.

        \item R$_{\rm \pm 3 \rm Mpc}$, R$_{\rm \pm 9 \rm Mpc}$ and R$_{\rm \pm 27 \rm Mpc}$ cubes and profiles: all the X-ray events, intrinsic and locally extrinsic, irrespective of whether they belong to the galaxy but within $3$ Mpc, $9$ Mpc, and $27$ Mpc of the source centre, are selected in 3D spherical comoving volumes. We construct the profiles and cubes to quantify the impact of contamination from the local vicinity on the intrinsic source emission. 
    \end{enumerate}

To quantify contributions from the emission associated with misclassified centrals, we use the \lstinline{SAT}$^{\rm sim}$ catalogue in the same stellar mass bin as the central galaxies. For galaxies in the three stellar mass bins $10^{10.5-11}\ \MSUN$, $10^{11-11.25}\ \MSUN$, and $10^{11.25-11.5}\ \MSUN$, we construct a total galaxy samples containing $N_{\rm tot}$ galaxies, of which there are $N_{\rm cen}$ centrals and $N_{\rm sat}$ satellites. The number of satellites is defined as $N_{\rm sat} = f_{\rm sat} N_{\rm tot}$, where $f_{\rm sat}$, the fraction of satellites, is varied at the values $ 0.01,\ 0.1,$ and $0.3$. The corresponding mean X-ray surface brightness profiles for a sample containing $ f_{\rm sat}$ satellites is
\begin{equation}
    \bar{\mathcal{S}}_{\rm X, tot}[r] =  \frac{\sum_{i=1}^{N_{\rm tot}} \mathcal{S}_{\rm X, tot}^{(i)}}{N_{\rm tot}}.
\end{equation}
The mean X-ray surface brightness profile components from the central galaxies, $\bar{\mathcal{S}}_{\rm X, cen}[r] $ and satellite galaxies,  $\bar{\mathcal{S}}_{\rm X, sat}[r]$ contributing to $\bar{\mathcal{S}}_{\rm X, tot}[r]$ are as follows:
\begin{multline}
    \bar{\mathcal{S}}_{\rm X, cen}[r] = (1 - f_{\rm sat}) \times \frac{\sum_{i=1}^{N_{\rm cen}}  \mathcal{S}_{\rm X, cen}^{(i)}}{N_{\rm cen}}\rm{\ and\ }\\ \bar{\mathcal{S}}_{\rm X, sat}[r]  = f_{\rm sat} \times \frac{\sum_{i=1}^{N_{\rm sat}}  \mathcal{S}_{\rm X, sat}^{(i)}}{N_{\rm sat}},\ \rm{respectively}.
\end{multline}
Therefore, $\bar{\mathcal{S}}_{\rm X, tot}[r] =  (1 - f_{\rm sat})\ \bar{\mathcal{S}}_{\rm X, cen}[r] +  f_{\rm sat}\ \bar{\mathcal{S}}_{\rm X, sat}[r]$.  We calculate the uncertainties in the total surface brightness profile of the total, central, and satellite components by bootstrapping.

\begin{figure}
    \centering
    \includegraphics[width=\linewidth]{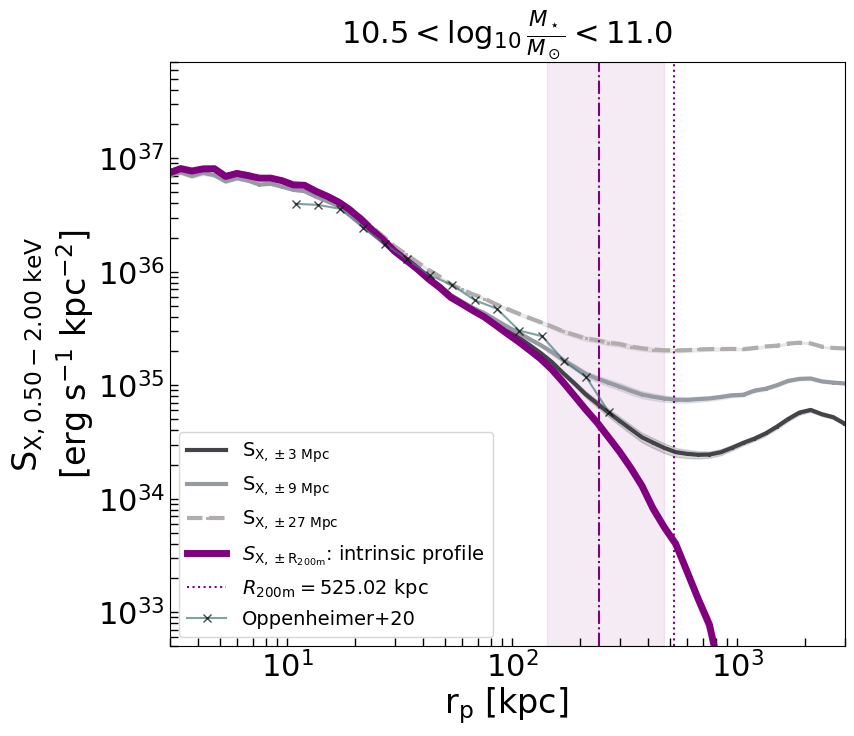}
    \caption{Mean X-ray surface brightness profiles in the stellar mass bin $M_{\star} = \big[ 10^{10.5}, 10^{11}\big]\ \MSUN$, corresponding to MW-like galaxies. The locally correlated large-scale structure contributions are shown by comparing the profiles obtained with photons selected within $\pm R_{200m}$ ({purple}) and those obtained within $\pm 3$ Mpc ({solid black}), $\pm 9$ Mpc ({solid grey}), $\pm 27$ Mpc ({dashed grey}) away from the halo centre. The crosses are from previous work by \cite{oppenheimer2020eagle}, where they generate mock X-ray observations using TNG-100. The vertical dashed line at $242$ kpc and the dotted line at $525$ kpc is the mean $R_{500c}$ and $R_{200m}$, respectively, of the $5,109$ galaxies used in the mass bin with the shaded area corresponding to the minimum and maximum $R_{500c}$ values. The shaded uncertainties on the profiles represent the variance obtained by bootstrapping.\label{fig:trans}}
\end{figure}

\section{Results}\label{sec:results}

Here, we present the main results of this work. In Sec.~\ref{subsec:intrinsic}, we discuss the outcome of fitting the analytical model to the stacked LC-TNG300 intrinsic profiles. In Sec.~\ref{subsec:lss},  we carry forward the current state-of-the-art to understand the impact of the locally correlated environment on the intrinsic X-ray surface brightness profile. Lastly, in Sec.~\ref{subsec:satboost}, we show the prominence of the effect due to misclassified centrals.

\subsection{Fitting analytic models to the intrinsic X-ray surface brightness profiles}
\label{subsec:intrinsic}

Fig.~\ref{fig:best-fit} and Fig.~\ref{fig:best-fit-halo} show the mean X-ray surface brightness profiles in three stellar mass and halo mass bins, respectively. We fit the surface brightness profiles with the analytic model introduced in Eq.~\ref{eqn:modified_beta} and provide the best-fit parameters in Tab.~\ref{tab:best-fit-params0} and Tab.~\ref{tab:best-fit-params1} for the stellar and halo mass bins, respectively. A simple $\beta$ model, described by Eq.~\ref{eqn:beta}, describes the lowest stellar and halo mass bins. However, the $\beta$ model does not describe the more massive stellar and halo mass bins. Therefore, we implement Eq.~\ref{eqn:modified_beta}, which very well describes the LC-TNG300 profiles in the stellar and halo mass bins across all masses.

The model has two break radii, the core radius $r_c$ and the scale radius $r_s$, where $r_c \ll r_s$. The scale radius $r_s$ affects the profile at radii beyond $ R_{\rm 500c}$, the mean $R_{\rm 500c}$ of the stacked mass bin. The model also introduces two slope parameters, $\beta$, which influences the X-ray surface brightness profile at radii $r<r_s$ and $\epsilon$, the slope that affects the profile at $r>r_s$. As $ R_{\rm 500c}$ is the radius used most commonly in observations, and $\beta$ quantifies the shape of the profile at radii $\leq  R_{\rm 500c}$, we discuss here the variations in $\beta$ across the stellar and halo mass bins.

The increasing stellar mass bins $10^{10.5-11}\ \MSUN$ , $10^{11.25-11.5}\ \MSUN$, and $10^{11.25-11.5}\ \MSUN$ result in decreasing values of $\beta$ from $0.4$, $0.28$, and $0.25$, respectively. The same trend holds from the halo mass bins, where increasing halo mass bins $10^{12.5-13}\ \MSUN$, $10^{13-13.5}\ \MSUN$, and $10^{13.5-14}\ \MSUN$ have decreasing values of $\beta$ from $0.57$, $0.30$, to $0.24$, respectively. One of the reasons for the decrease is attributed to the mass-dependence of the feedback prescriptions in the TNG300 model, i.e., stellar mode dominating at lower masses, and the kinetic and thermal modes of energy injection by AGN dominating galaxies with $M_\star\geq 10^{10.5}\ \MSUN$~\citep{weinberger2016simulating}. As discussed in \cite{weinberger2018supermassive}, the two AGN-related feedback channels depend on the black-hole (BH) accretion rates. For the high accretion rates, the thermal mode causes the gas cells close to the galactic centre to heat, eventually releasing the energy radiatively in X-rays. However, for lower accretion rates, the kinetic mode kicks in, depositing energy through winds and jets in random directions away from the BH. This causes gas heating and, hence, X-ray emission via cooling at a larger distance away from the galaxy centre, overall causing a flattening of the radial X-ray surface brightness profile as shown by the slope $\beta$ in Fig.~\ref{fig:best-fit} and~\ref{fig:best-fit-halo}. This reasoning is further consolidated with studies detailing the effects of the kinetic mode of BH feedback in the TNG model \citep{terrazas2020relationship}, its impact on the gas properties, i.e., temperature, entropy, density, and CGM fraction~\citep{Zinger2020tnggas, Davies2020feedback}, and its correlation with X-ray emission \citep{truong2020x, oppenheimer2020eagle, Truong2021MbhXrays, Ayromlou2023closure}. \cite{sorini2024impact} show a similar trend of decreasing slope for the gas density profiles with the SIMBA~\citep{dave2019simba} suite of simulations. With simulation comparison projects like \textsc{camels} simulations~\citep{Navarro2021camel}, it is possible to study the correlation between X-ray emission around galaxies with different feedback implementations, the exploration of which we leave for a future study. 

In the following two sections, we focus on how the projection effects result in deviations from the intrinsic profile. We discuss the effects on the stellar mass bins as it is an observationally available mass proxy.

\subsection{Locally correlated environment}
\label{subsec:lss}

For the MW-mass bin, $M_{\star} = 10^{10.5-11}\ \MSUN$, with mean $ R_{\rm 200m}  =525$ kpc, we generate cubes within $\pm 3 \rm \ Mpc$, $\pm 9 \rm\  Mpc$ and $\rm \pm 27\  \rm Mpc$ from the galaxy center. This corresponds to photons selected within $\sim 5.7 \times R_{\rm 200m}$, $\sim 17 \times R_{\rm 200m}$, and $\sim 51.4 \times R_{\rm 200m}$, respectively. Fig.~\ref{fig:trans} presents the impact of the locally correlated LSS for MW-like halos. We show the intrinsic profile that is obtained by selecting events within $\pm R_{200m}$ of the galaxy centre in purple. We compare our result with previous work from \cite{oppenheimer2020eagle}, shown with the crosses in Fig.~\ref{fig:trans}. They predict the profile between $10-242$ kpc for a stellar mass sample in the mass range $10^{8.2-11.39} \MSUN$ containing $\sim 400$ galaxies. Their stellar mass sample is divided into low sSFR and high sSFR. We take the mean of the low sSFR and high sSFR profiles generated in their work and compare them with ours. The halo mass range corresponding to their stellar mass bin is $10^{\sim 12.3-13}\ \MSUN$, which entails the mean halo mass $ M_{\rm 200m} = 10^{12.7} \ \MSUN$, same as that of the MW mass bin used in our work. \cite{oppenheimer2020eagle} use the TNG-100 simulation and synthesize mock X-ray observations for individual halos as opposed to this work that uses TNG300 and synthesizes mock X-ray observations within the lightcone. Despite these differences, this work's predicted X-ray intrinsic emission profiles are in good agreement with \cite{oppenheimer2020eagle}.

We find that increasing the volume by $\frac{4 \pi}{3}(5.7^3 - 1^3)R^3_{200m} = 771.5 \times R^3_{200m}$, i.e., by including events in $\pm 3$ Mpc, boosts the X-ray surface brightness profile at $R_{200m}$ by a factor of 5.2. We show this with the thick black line in Fig.~\ref{fig:trans}. We find deviations from the true intrinsic profile due to events selected within $\pm 3$ Mpc at $\sim 150$ kpc, which corresponds to $\approx0.6\times  R_{500c}$ and $\approx 0.3\times  R_{200m}$ for the MW-stellar mass bin. 

When considering the events in $\pm9$ Mpc, i.e., increasing the volume used to compute the X-ray surface brightness profiles by $\frac{4 \pi}{3}(17^3 - 1^3)R^3_{200m} = (2.0\times 10^4)\ R^3_{200m}$, we find that the X-ray surface brightness profile at $R_{200m}$ is boosted by a factor of 16.8 This is shown by the grey curve in Fig.~\ref{fig:trans}. We find deviations from the true intrinsic profile due to events selected within $\pm 9$ Mpc, at $\sim 100$ kpc, which corresponds to $\approx0.4\times  R_{500c}$ and $\approx 0.2\times  R_{200m}$ for the MW-stellar mass bin. 

For events selected in $\pm27$ Mpc or by including events in a volume of $(8.2\times 10^4)\ R^3_{200m}$, the X-ray surface brightness profile at $R_{200m}$ is boosted by $47.3$. This is shown by the light grey dashed curve in Fig.~\ref{fig:trans}. The profile, in this case, remains unchanged only at radii $\leq 40$ kpc, which corresponds to  $\approx0.2\times  R_{500c}$ and $\approx 0.08\times  R_{200m}$ for the MW-stellar mass bin. 

In conclusion, we show for the first time the effect of the local environment on the true intrinsic profile of a mean MW-like stacked X-ray surface brightness profile. Namely, we find deviations from the true profile at $\approx 0.3\times  R_{200m}$, $\approx 0.2\times  R_{200m}$, and $\approx 0.08\times  R_{200m}$ by including events out to $\sim 5.7 \times R_{\rm 200m}$, $\sim 17 \times R_{\rm 200m}$, and $\sim 51.4 \times R_{\rm 200m}$, respectively. The black, grey, and dashed-grey curves in Fig.~\ref{fig:trans} show that increasing the integration volume swamps the features of the intrinsic profile at radii closer to the galaxy centre.

In observations, one is sensitive to the complete line of sight towards the observer. Therefore, this effect can be corrected by subtracting a background emission level determined empirically from the observed surface brightness at a large distance from the halos of interest (assuming spherical symmetry in the large-scale emission). We test the impact of a simple, conventional background subtraction, where we subtract the mean value of the surface brightness profile beyond the $R_{\rm 200m}$ (vertical purple dotted line) from the respective grey, and dashed-grey curves in Fig.~\ref{fig:trans}, as is most commonly done in observations. We find that the resulting profiles after background subtraction agree well with the intrinsic profile (solid purple line) out to $R_{\rm 500c}$ (the vertical dashed-dotted line). Beyond $R_{\rm 500c}$, such a subtraction underestimates the X-ray emission at the outskirts. However, this must be further tested with larger-volume lightcones than used here (see further discussion in Sec.~\ref{sucsec:lss_discussion}). Nevertheless, given the setup used in this work, we can, for the first time, probe this effect locally around the halo and show its impact in stacking experiments.

\begin{figure*}

    \centering
    \begin{minipage}{0.34\textwidth}
        \includegraphics[width=\linewidth]{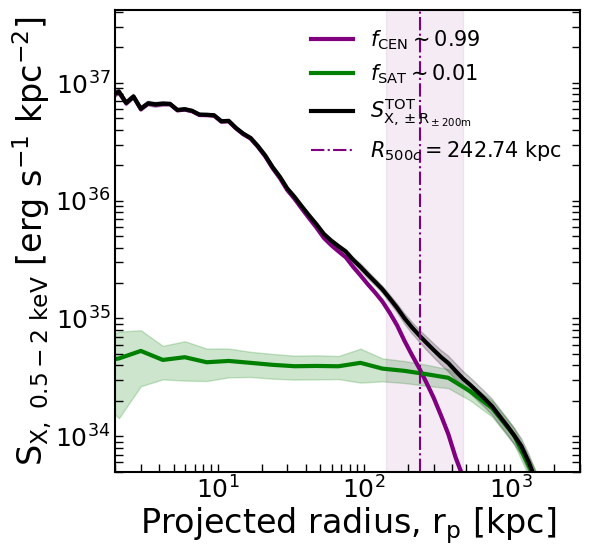}
    \end{minipage}%
    \centering
    \begin{minipage}{0.33\textwidth}
        \includegraphics[width=\linewidth]{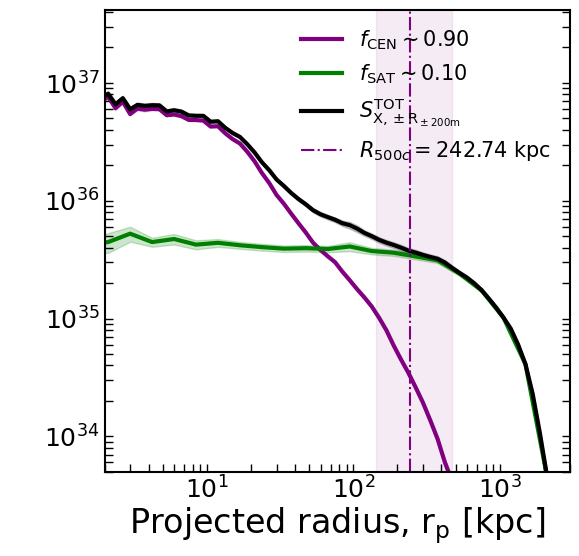}
    \end{minipage}%
    \centering
    \begin{minipage}{0.33\textwidth}
        \includegraphics[width=\linewidth]{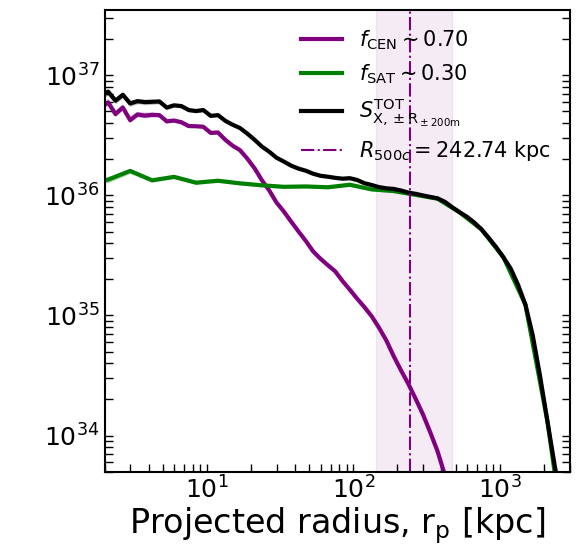}
    \end{minipage}%
    \caption{Effect of centrals ({purple}) and misclassified-centrals ({green}), in the stellar mass bin $M_{\star} = \big[ 10^{10.5}, 10^{11}\big]\ \MSUN$, on the total X-ray surface brightness profiles ({black}). The total sample is constructed such that there are $1\%$ satellites and $99\%$ centrals ({left panel}), $10\%$ satellites and $90\%$ centrals ({middle panel}), and $30\%$ satellites and $70\%$ centrals ({right panel}). The vertical dashed line at $242$ kpc is the mean $R_{500c}$ of the $5109$ galaxies used in the mass bin with the shaded area corresponding to the minimum and maximum values. The shaded regions on the profiles are the uncertainties obtained by bootstrapping. \label{fig:censat}}
\end{figure*}

\begin{figure*}

    \centering
    \begin{minipage}{0.34\textwidth}
        \includegraphics[width=\linewidth]{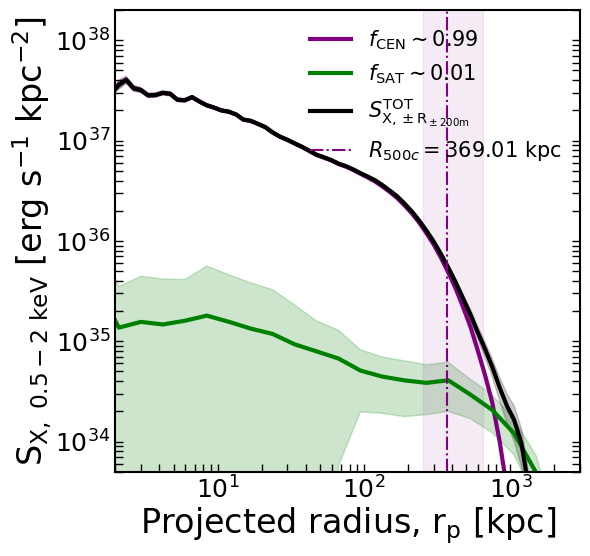}
    \end{minipage}%
    \centering
    \begin{minipage}{0.33\textwidth}
        \includegraphics[width=\linewidth]{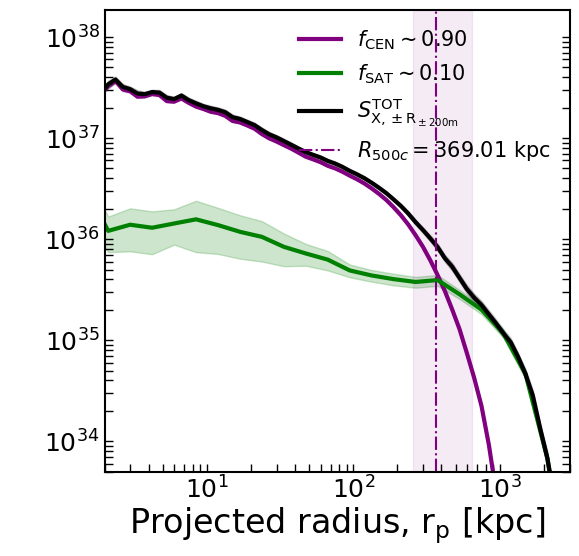}
    \end{minipage}%
    \centering
    \begin{minipage}{0.33\textwidth}
        \includegraphics[width=\linewidth]{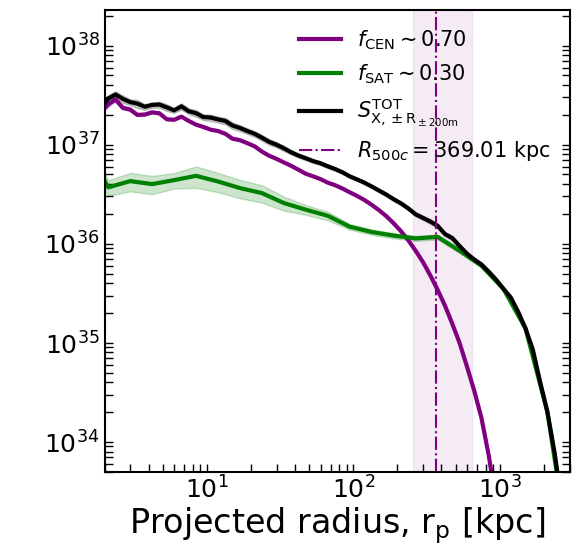}
    \end{minipage}%
    \caption{Effect of centrals ({purple}) and misclassified-centrals ({green}), in the stellar mass bin $M_{\star} = \big[ 10^{11}, 10^{11.25}\big]\ \MSUN$, on the total X-ray surface brightness profiles ({black}). The total sample is constructed with $1\%$  ({left panel}), $10\%$  ({middle panel}), and $30\%$  ({right panel}) satellites. The vertical dashed line at $369$ kpc is the mean $R_{500c}$ of the $680$ galaxies used in the mass bin with the shaded area corresponding to the minimum and maximum values. \label{fig:censat1}}
\end{figure*}

\begin{figure*}

    \centering
    \begin{minipage}{0.35\textwidth}
        \includegraphics[width=\linewidth]{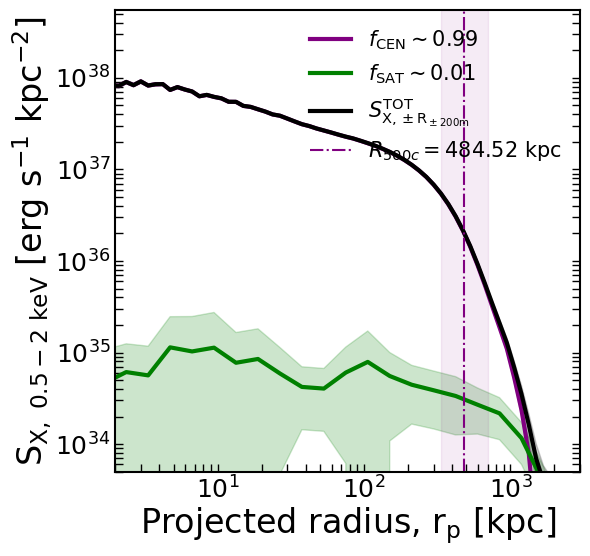}
    \end{minipage}%
    \centering
    \begin{minipage}{0.32\textwidth}
        \includegraphics[width=\linewidth]{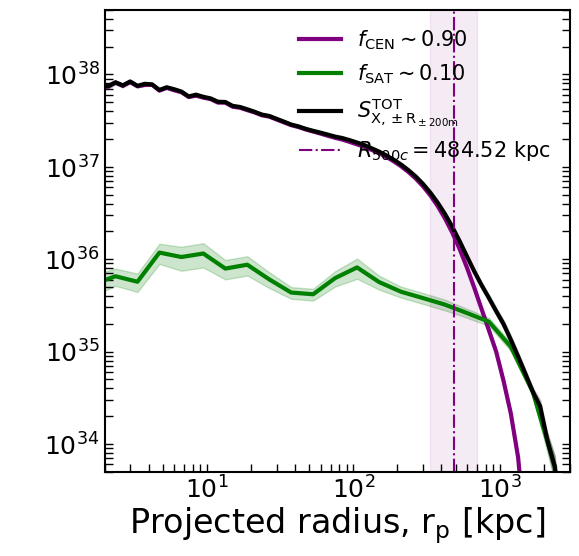}
    \end{minipage}%
    \centering
    \begin{minipage}{0.32\textwidth}
        \includegraphics[width=\linewidth]{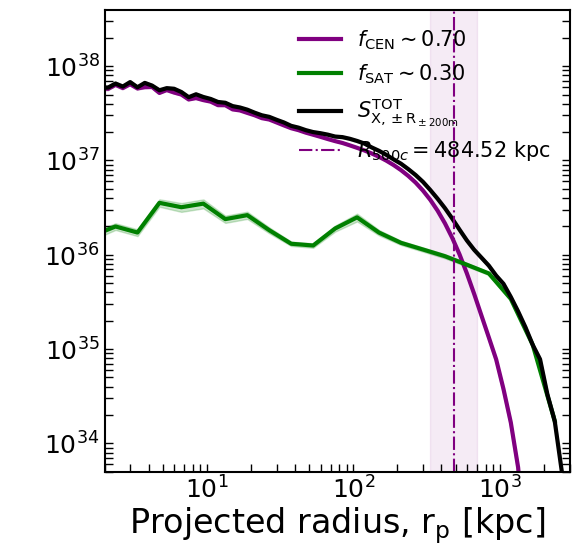}
    \end{minipage}%
    \caption{Effect of centrals ({purple}) and misclassified-centrals ({green}), in the stellar mass bin $M_{\star} = \big[ 10^{11.25}, 10^{11.5}\big]\ \MSUN$, on the total X-ray surface brightness profiles ({black}). The total sample is constructed with $1\%$  ({left panel}), $10\%$  ({middle panel}), and $30\%$  ({right panel}) satellites. The vertical dashed line at $484.5$ kpc is the mean $R_{500c}$ of the $305$ galaxies used in the mass bin with the shaded area corresponding to the minimum and maximum values.\label{fig:censat2}}
\end{figure*}

\subsection{The effect of misclassified centrals}
\label{subsec:satboost}

In simulations, we completely and accurately classify central and satellite galaxies within a given stellar mass bin. Using this to our advantage, here, we make a precise prediction of the average emission arising from stacking around satellite galaxies by considering the following fraction of satellite (or misclassified centrals) contaminating the total galaxy sample $0.01$, $0.1$ and $0.3$. We select these contamination fractions of satellites by bootstrapping over the entire satellite galaxy sample. Tab.~\ref{tab:stellar_mass_bins} details the number of centrals and satellites in the three stellar mass bins presented here. We show our findings in Fig.~\ref{fig:censat},~\ref{fig:censat1}, and~\ref{fig:censat2}, corresponding to stellar mass bins $10^{10.5-11}\ \MSUN$, $10^{11-11.25}\ \MSUN$, and $10^{11.25-11.5}\ \MSUN$, respectively. 

Fig.~\ref{fig:censat} quantifies the effect of misclassified centrals on the total surface brightness profile, shown with the solid black line, in the MW-stellar mass bin of $10^{10.5-11}\ \MSUN$. We show the integrated profile obtained by stacking only the satellite galaxies misclassified as centrals in green and the intrinsic profile due to central galaxies in purple. The point at which the profile due to misclassified centrals ({green}) intersects the intrinsic central galaxy ({purple}) profile represents the point at which the effect due to misclassified centrals contributes $\sim 50\%$ to the total (centrals and satellites) emission. For a sample with $1\%$ ({left panel}) and $10\%$ ({central panel}) contaminating satellites, the effect due to misclassified centrals dominates over the intrinsic central galaxy emission at radii {$\geq 1.04 \times R_{500c}$ ($\sim 252$ kpc)} and {$\geq 0.24 \times R_{500c}$ ($\sim 59$ kpc)}, respectively. In a sample with $30\%$ satellites {contaminants} ({right panel}), the effect due to misclassified centrals dominates over the intrinsic central galaxy emission at radii {$\geq 0.11 \times R_{500c}$ ($\sim 27$ kpc)}. 

Analogously, Fig.~\ref{fig:censat1} quantifies the effect of misclassified centrals on the total surface brightness profile, shown with the solid black line, for galaxies in the stellar mass bin $10^{11-11.25}\ \MSUN$. For a sample with $1\%$ ({left panel}) contaminating satellites, we find that the effect of misclassified centrals has negligible impact within radii $\leq  R_{500c}$. For $10\%$ satellite contamination ({central panel}), the effect due to misclassified centrals dominates over the intrinsic central galaxy emission at radii {$\geq 1.1 \times R_{500c}$} {($\sim 400$ kpc)} and for $30\% $ satellites contamination ({right panel}), the effect due to misclassified centrals dominates over the intrinsic central galaxy emission at radii {$\geq 0.63 \times R_{500c}$ ($\sim 231$ kpc)}.

Finally, Fig.~\ref{fig:censat2} quantifies this effect for the largest mass bin we are considering, $10^{11.25-11.5}\ \MSUN$. For a sample with $1\%$ ({left panel}), $10\%$ ({centre panel}) and $30\%$ satellite contamination ({right panel}), the effect due to misclassified centrals remains negligible at all radii $\leq  R_{500c}$. More precisely, the $1\%$ satellite contamination has a negligible impact on the total surface brightness profile. The $10\%$ and  $30\%$ satellite contamination dominate over the intrinsic central galaxy profile at {$\geq 1.7 \times R_{500c}$ ($\sim 804$ kpc) and $\geq 1.2 \times R_{500c}$ ($\sim 565$ kpc)}, respectively.

As expected, we find from Fig.~\ref{fig:censat},~\ref{fig:censat1}, and~\ref{fig:censat2} that the effect due to misclassified centrals becomes increasingly important as we probe lower stellar mass bins in stacking experiments. As we go to higher mass bins, the decreasing impact of the effect of misclassified centrals is attributed to the satellites - of the same stellar mass bin - residing in less massive parent halos (see further explanation in Sec.~\ref{sec:disandcon}).

Fig.~\ref{fig:fsatintersect} presents the radial fraction of the X-ray emission from the intrinsic CGM - for four different levels of misclassified centrals contamination - in three stellar mass bins $10^{10.5-11}\ \MSUN$, $10^{11-11.25}\ \MSUN$, and $10^{11.25-11.5}\ \MSUN$, respectively. We complement the conclusions from Fig.~\ref{fig:censat}, \ref{fig:censat1}, and \ref{fig:censat2} by showing that for MW-like galaxies, the effect of misclassified centrals at the lowest contamination fraction of $0.01$ results in the intrinsic central galaxy emission contributing only {$\sim 51\%$} of the total emission at $R_{500c}$. This further deteriorates with the increasing fraction of misclassified centrals in the galaxy sample, where fractions of $0.1$, $0.3$, or $0.5$ result in the intrinsic emission contributing {$\sim 9\%$, $\sim 3\%$ and $\sim 1\%$} of the total emission at $R_{500c}$, respectively. This effect is less pronounced for the intermediate and most massive stellar mass bins of $10^{11-11.25}\ \MSUN$ and $10^{11.25-11.5}\ \MSUN$ compared to the MW-like stellar mass bin. In the case of the $10^{11-11.25}\ \MSUN$ mass bin, for a fraction of misclassified centrals of $0.01$, $0.1$, $0.3$, or $0.5$, the central intrinsic emission contributes $\sim 93\%$, {$\sim 55\%$, $\sim 23\%$ and $\sim 11\%$} of the total emission at $R_{500c}$, respectively. Analogously, for the stellar mass bin of $10^{11.25-11.5}\ \MSUN$, for a fraction of misclassified centrals of $0.01$, $0.1$, $0.3$, or $0.5$, the central intrinsic emission contributes $\sim 99\%$, $\sim 86\%$, {$\sim 61\%$} and {$\sim 40\%$} of the total emission at $R_{500c}$, respectively. Therefore, we present a clear trend of the increasing importance of the effect of misclassified centrals, not only due to the increasing fraction of satellite contamination in the galaxy sample but also due to the decreasing stellar mass bins.

\section{Discussion}
\label{sec:disandcon}

This work uses a TNG300-based forward model for the hot CGM emission and presents, for the first time, the effect of locally correlated large-scale structure around a halo, the effect of misclassified centrals in stacked hot CGM galactocentric profiles and the effect of the choice of centre. Our findings are vital in light of our explorations for studying X-ray emission around lower mass galaxies with stacking experiments. We divide the discussion of our results by focusing on the projection effects due to the locally correlated environment, first, in Sec.~\ref{sucsec:lss_discussion}, second, the effect of misclassified centrals in Sec.~\ref{subsec:censat_discussion}, and lastly, the effect of miscentering in Sec.~\ref{subsec:xray2true_cen}.

\subsection{Locally correlated environment}
\label{sucsec:lss_discussion}

The locally correlated environment for MW-like stellar mass galaxies boosts the mean galactocentric X-ray emission at radii $\leq R_{200m}$. More precisely, at $R_{200m}$ the X-ray emission is boosted $5.2\times$, $16.8\times$ and $47.3\times$ by including emission $\pm 3$ Mpc, $\pm 9$ Mpc, $\pm 27$ Mpc away from the halo centre, respectively. 

Increasing the volume over which the intrinsic galactocentric profiles are measured leads to increased X-ray emission at smaller radii. This increased X-ray emission, in projection, is attributed to the local environment in which the galaxy resides. The upturn in the black line in Fig.~\ref{fig:trans}, corresponding to including events within $\pm 3$ Mpc ($\sim 5.7 × R_{200m}$) of the galaxy centre, signifies the presence of other X-ray emitting halos $\gtrsim 1$ Mpc away from MW-like galaxies, i.e., $\approx 2\times  R_{200m}$ or $\approx 4\times  R_{500c}$. This "upturn" feature is washed out as we start including events in even larger volumes, as shown by grey and dashed-grey lines (corresponding to events selected within $\pm 9$ Mpc and $\pm 27$ Mpc, respectively) in Fig.~\ref{fig:trans}. This supports our finding that an observer starts probing the averaged hot gas emission from all galaxies in projection along the line of sight, drowning out the features due to the local environment when integrating events over larger volumes around a galaxy centre.

Further improvements to study this effect would involve including events to even larger radii around the galaxy. Given the limited area of the lightcone, $47.28\rm \ deg^2$, we are constrained in studying this effect out to $\pm 27$ Mpc. To quantify the effects of the locally correlated large-scale environment to even larger distances than those explored here, we need larger-volume lightcones, which is possible with larger cosmological hydrodynamical simulations like Magneticum~\citep{dolag2015magneticum}, MillenniumTNG~\citep{pakmor2023millenniumtng, hernandez2023millenniumtng}, and FLAMINGO~\citep{Schaye:2023aa}. Additionally, future studies could study how different simulation feedback prescriptions can impact the trend observed in the X-ray surface brightness profiles by including events in larger volumes. It is particularly interesting to understand how the physics implemented in hydrodynamical simulations impacts the baryon spread around galaxies, i.e., the radius at which all the emissions converge to the mean background level in X-rays. We leave the study of these aspects to future work, which is made possible with the advent of projects like the \textsc{camels} simulations~\citep{Navarro2021camel}; see, e.g., \cite{gebhardt2024cosmological} and \cite{sorini2022baryons}.

\begin{figure*}

    \centering
    \begin{minipage}{\textwidth}
        \includegraphics[width=\linewidth]{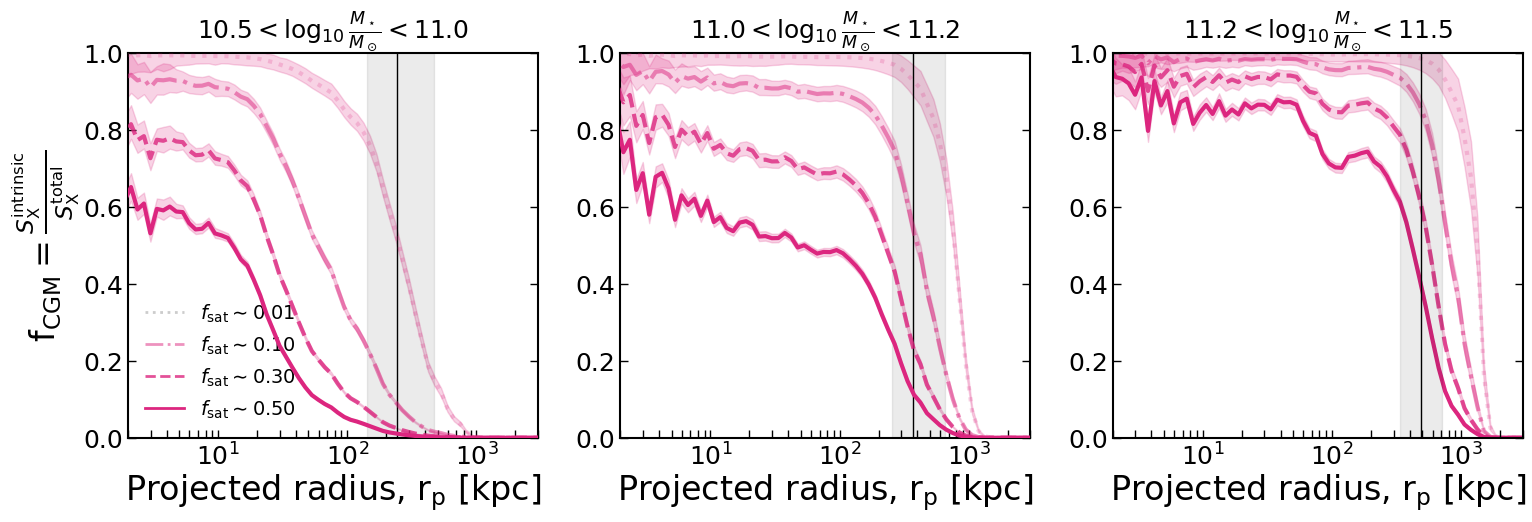}
    \end{minipage}%
    \caption{Fraction of X-ray emission from the intrinsic hot CGM for different levels of satellite contamination (misclassified centrals) in the total galaxy sample. The fraction of X-ray emission from the intrinsic hot CGM for misclassified central contamination fractions of $0.01$ (dotted line), $0.1$ ({dash-dotted line}), $0.3$ ({dashed line}), and $0.5$ ({solid line}) are shown for three different mass bins: MW-like galaxies with stellar-masses in $10^{10.5-11}\ \MSUN$ ({left panel}),  $10^{11-11.25}\ \MSUN$ ({middle panel}), and $10^{11.25-11.5}\ \MSUN$ ({right panel}). The shaded regions around each curve correspond to the uncertainty of the mean profile obtained by bootstrapping. The vertical black solid line at $242$ kpc in the left panel, $369$ kpc in the middle panel, and $484.5$ kpc in the right panel corresponds to the mean $R_{500c}$ of the respective mass bins with the shaded area signifying the minimum and maximum values.  We find that the contaminating effect of the misclassified centrals on the stacked profiles increases when the stellar mass decreases.\label{fig:fsatintersect}}
\end{figure*}

\subsection{The effect of misclassified centrals}
\label{subsec:censat_discussion}

We quantify the impact of the misclassified centrals (satellite-boost) residing in the same stellar mass bin as the central galaxies in Fig.~\ref{fig:censat},~\ref{fig:censat1}, and~\ref{fig:censat2} for the stellar mass bins  $10^{10.5-11}$,  $10^{11-11.25}$, and  $10^{11.25-11.5}$, respectively. For satellite galaxies with stellar masses of MW-like galaxies, the emission around satellites dominates over central emission because the satellites of the stellar mass range $10^{10.5-11} \MSUN$ reside in distinct halos with masses $M_{\rm 200m} = 10^{12.3-15.1}$, corresponding to a mean halo mass {$M_{200m}  = 10^{13.7}\ \MSUN$}. We find that the satellite subhalos probe the emission of their parent halos, contributing to the stack with off-centred surface brightness profiles of their more massive parent halo. This contribution causes shallower slopes for the total emission (central and satellites) in the MW-stellar mass bin. In cases where $30\%$, $10\%$, or $1\%$ of the satellites contribute to the total emission, we have shown that they dominate the measured total profile at radii $\geq 0.11 \times R_{500c}$, $\geq 0.24 \times R_{500c}, $ and $\geq 1.04 \times R_{500c}$, respectively. We further investigate the effect of misclassified centrals in the more massive stellar mass bins of  $10^{11-11.25}\ \MSUN$, and $10^{11.25-11.5}\ \MSUN$ in Fig.~\ref{fig:censat1} and Fig.~\ref{fig:censat2}, respectively. We find the effect of misclassified centrals, with contamination fractions of $0.01$, $0.1$, and $0.3$, less affects the $10^{11-11.25}\ \MSUN$, and $10^{11.25-11.5}\ \MSUN$ mass bins as opposed to the MW-stellar mass bin. We reaffirm this with the trend shown in Fig.~\ref{fig:fsatintersect}, where the fraction of intrinsic emission at the mean $R_{500c}$ is lowest for the MW stellar mass bin for all contamination levels, as opposed to the more massive stellar mass bins.

To understand why the MW-stellar mass bin is most affected by the effect of misclassified centrals, we first look at the {host} halo masses of the satellites in the higher stellar mass bins of $10^{11-11.25}\ \MSUN$ and $10^{11.25-11.5}\ \MSUN$. These satellites reside in distinct halos of {mean $M_{\rm 200m}$} {$\sim 10^{14}\ \MSUN$ and $\sim 10^{14.2}\ \MSUN$}, respectively. This is a direct implication arising from the difference in the stellar-to-halo-mass relation for central and satellite galaxies~\citep{shuntov2022cosmos2020}. Given that the mean halo mass of the central galaxies in the LC-TNG300 stellar mass bins of $10^{11-11.25}\ \MSUN$, and $10^{11.25-11.5}\ \MSUN$ is $10^{13.3}\ \MSUN$ and $10^{13.6}\ \MSUN$, respectively, we require large fraction of contamination to see the effect of misclassified centrals compared to that for the MW-stellar mass bin. More precisely, we find that the effect of misclassified centrals dominates over the central galaxy emission at $ R_{500c}$ with satellite contamination fractions, $f_{\rm cont,\ sat}$, of $0.01, 0.14$, and $0.77$ for the stellar mass bins $10^{10.5-11} \ \MSUN$, $10^{11-11.25} \ \MSUN$, and $10^{11.25-11.5} \ \MSUN$, respectively. Hence, the contamination fractions probed here ($f_{\rm cont,\ sat}= 0.01, 0.1, $ and $0.3$), which are closer to the realistic $f_{\rm cont,\ sat}$ in observations, do not significantly affect the stellar mass bins $10^{11-11.25}$, and $10^{11.25-11.5}$, respectively.

\begin{figure}
    \centering
    \includegraphics[width=.9\linewidth]{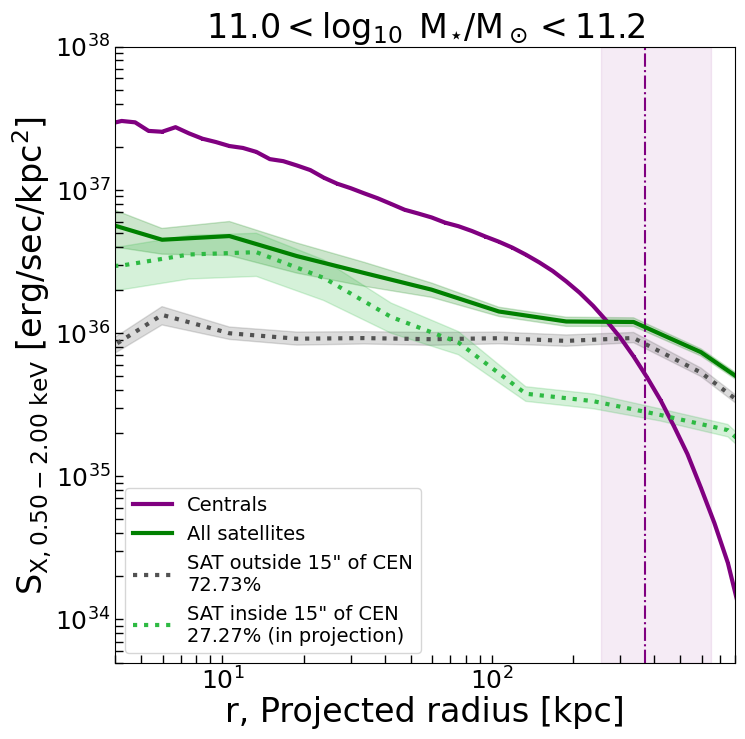}
    \caption{Effect of further selection on the satellites misclassified as central galaxies on the X-ray surface brightness profile. The total population of satellites misclassified as centrals ({solid green line}) is divided into satellites within 15" of a central ({green dotted line}) and satellites farther than 15" from a central ({grey dotted line}) in projection on the sky. We show that satellites (misclassified as centrals) within 15” of a central galaxy have a steeper profile than the case where we consider all the satellites, thereby reducing the overall flattening of the X-ray surface brightness profile due to misclassified centrals. Contrarily, the satellites farther than 15” from a central have a constant surface brightness profile. The central galaxy profile is shown in purple. The vertical dashed line at $369$ kpc is the mean $R_{\rm 500c}$ of the $680$ galaxies used in the M31 stellar mass bin, and the shaded area corresponds to the minimum and maximum $R_{\rm 500c}$  value of the central galaxies in the M31-mass bin. The shaded uncertainties on the profiles represent the variance obtained by bootstrapping.
    \label{fig:sat_in_15arcsec}
    }
\end{figure}

We now discuss how to observationally mitigate the effect of satellites misclassified as centrals using X-ray information. We investigate this issue here under the assumption that the observational galaxy sample is accurately classified into centrals and satellites; in such a case, the effect of misclassified centrals is dominated by instrument resolution, i.e., only satellite objects that are farther from the central than the distance corresponding to the instrument resolution can be easily disregarded as misidentified centrals. However, we must also note that using the X-ray information to mitigate the effect of satellites (based on their proximity to a central) requires the individual halos to be detected in X-rays, which, given the faintness of the CGM emission, is currently not viable for MW-mass halos with survey instruments like eROSITA (see further discussion in Sec.~\ref{subsec:xray2true_cen}). For more massive (M31-like) halos, which have higher chances of detection in X-rays, we study how such segregation of misclassified centrals based on their proximity to a central in the sky would impact the mean X-ray surface brightness profile from misclassified centrals. We find that the mean X-ray profile from M31-mass satellite galaxies within $15"$ of a central galaxy drops by $25\%$ at $R_{500c}$ as opposed to the case where we consider all the satellite galaxies as misclassified centrals, as shown in Fig.~\ref{fig:sat_in_15arcsec}. Therefore, such a technique for mitigating satellite contamination reduces the flattening of the mean X-ray surface brightness profile from satellites. However, an important caveat when selecting centrals without satellites in the vicinity is that such an experiment would give a biased result in the detected CGM itself, where one would be probing the CGM of galaxies only in isolated environments.

Given these findings, we conclude that the contribution of satellites in MW-like galaxy samples containing satellites must not be neglected in stacking analysis. Observationally, to disentangle the contamination of the misclassified centrals, we must fit the profiles contributing to the intrinsic central galaxy jointly with the component capturing the effect of misclassified centrals for a given fraction of satellite contamination. 

\subsection{The effect of miscentering: X-ray centre offset from the centre defined by the minimum of the dark matter potential}
\label{subsec:xray2true_cen}

Here, we study the effect of offset between the observationally defined X-ray centre and the centre defined by the minimum of the halo dark matter potential. The X-ray centre is computed using a luminosity-weighted average of all the counts within the spherical $R_{\rm 200m}$ cubes of the central halo. In this work, we use the input centre or the true centre as the location of the most bound particle within the halo as given by the FoF/\textsc{subfind} algorithm applied on TNG300 to obtain halo catalogues. We use the projected positions on the sky for computing the offset, $\Delta_{\rm X-ray - input}$, and convert these offsets to physical scales using the redshift of the halo. The mean of the $\Delta_{\rm X-ray - input}$ distribution is $70$ kpc, with the $16^{\rm th}-84^{\rm th}$ percentiles ranging between $13 - 131$ kpc, which is consistent with other works in literature~\citep{Seppi2023offset, popesso2024perils}. The corresponding mean value of the $M_{\rm 200m}$ distribution is at $10^{13.3} \MSUN$, i.e., within the M31-mass bin, with $16^{\rm th}-84^{\rm th}$ percentiles ranging between $10^{12.4 - 13.5} \MSUN$. The mean offset in the three halo mass bins considered in this work, $10^{12.5-13} \MSUN$,  $10^{13-13.5} \MSUN$,  and $10^{13.5-14} \MSUN$, are $52$ kpc, $66$ kpc, and $79$ kpc, respectively; thereby, showing a correlation of increasing X-ray to true centre offset with increasing halo masses.

A galaxy stacking experiment with an eROSITA-like instrument that uses the dark matter halo potential minima as the centre would be affected by this miscentering effect. Assuming that the X-ray halo is detected, the objects for which the miscentering effect can be corrected are those where the offset between the X-ray and true centre is $\gtrsim 30"$, the resolution of eROSITA. In our lightcone, we report that $\sim 13\%$ of objects have resolvable offsets (i.e., the offset separation $\gtrsim 30"$), which, if classified in $M_{\rm 200m}$ halo mass bins are as follows. For the $10^{12.5-13} \MSUN$ and $10^{13-13.5} \MSUN$ mass bins, $13\%$ of objects have resolvable offsets; however, for the more massive $10^{13.5-14} \MSUN$ mass bin, $17\%$ of objects have resolvable offsets. However, we must note that individual MW mass halos in X-rays with eROSITA are below the detection threshold. Therefore, detecting their X-ray peak is currently unfeasible, but might be possible with future observatories.

\section{Summary}
\label{sec:summary}

The main conclusions from this work are summarised as follows.
\begin{enumerate}
    
    \item We present an analytical model (Eq.~\ref{eqn:modified_beta}) that well-describes the intrinsic $S_{X}$-profile in LC-TNG300 across the stellar mass bins $10^{10.5-11} \MSUN$,  $10^{11-11.25} \MSUN$,  and $10^{11.25-11.5} \MSUN$ and halo mass bins of $10^{12.5-13} \MSUN$,  $10^{13-13.5} \MSUN$,  and $10^{13.5-14} \MSUN$. We provide the best-fitting parameters for the analytical model in Tab.~\ref{tab:best-fit-params0} and Tab.~\ref{tab:best-fit-params1} for the stellar and halo mass bins explored in this work.
    
    \item We carry forward the current state-of-the-art modelling analysis presented in \cite{oppenheimer2020eagle} by also showing the impact of the locally correlated environment on the measured X-ray surface brightness profiles.
    
    \item We present, for the first time, the effect of misclassified centrals in stacking experiments for three stellar mass bins  $10^{10.5-11} \MSUN$,  $10^{11-11.25} \MSUN$,  and $10^{11.25-11.5} \MSUN$. We find that the contaminating effect of the misclassified centrals on the stacked profiles increases when the stellar mass decreases.
    
    \item For the MW-like galaxies, we conclude that the contribution of satellites (or misclassified centrals) can not be neglected in stacking analysis (see Fig.~\ref{fig:fsatintersect}). In cases where $30\%$, $10\%$, or $1\%$ of the satellites contribute to the total emission of MW-like galaxies, we have shown that they dominate the measured total $S_X$ profile at radii $\geq 0.11 \times R_{500c}$, $\geq 0.24 \times R_{500c}, $ and $\geq 1.04 \times R_{500c}$, respectively.

    \item {We report the mean offset between the centre of the halo defined using the X-ray peak, and that obtained from the minimum of the halo potential is $70$ kpc. }
    
\end{enumerate}

Modelling observed CGM profiles and comparing them with simulations is crucial to constrain the different galaxy formation models and to understand how the feedback and physics prescriptions affect the hot CGM profile. Current state-of-the-art cosmological hydrodynamical simulations are calibrated on the stellar mass function and successfully reproduce realistic galaxy populations. Despite this, \cite{davies2020quenching} show that EAGLE and IllustrisTNG predict different total gas mass fractions, affecting the observed CGM properties at MW-masses. Similarly, \cite{khrykin2024cosmic} shows how the hot gas is sensitive to the different feedback variants within SIMBA. The model and methodology presented here provide the machinery to compare the measured hot CGM profiles among current simulations in future works. 

Future X-ray missions, on the observation side, like Athena~\citep{nandra2013hot}, AXIS~\citep{mushotzky2019advanced}, HUBS~\citep{cui2020hubs} will push our current detection limits and provide us with (i) the spatial resolution to reach higher redshifts, (ii) better quantify point-like source contamination within the hot CGM, (iii) the spectral resolution to allow disentangling components via spectral fitting and (iv) the grasp to reach even fainter surface brightness levels. This work, focusing on modelling the projection effects, is a step towards exploiting the information provided by the next-generation telescopes to better understand the intrinsic hot CGM emission. The prospects of this framework would be to extend it to interpret hot X-ray CGM measurements in stacking experiments by accounting for all projection effects, i.e., not only the impact due to the local environment of the halo and the effect due to misclassified centrals but also the emission from other X-ray sources such as AGN and XRB.

\begin{acknowledgements}
SS would like to thank Fulvio Ferlito, Matteo Guardiani, Christian Garrel, Emre Bahar, Jeremy Sanders, and the anonymous referee for all the useful scientific discussions.

G.P. acknowledges financial support from the European Research Council (ERC) under the European Union's Horizon 2020 research and innovation program "Hot Milk" (grant agreement No. 865637) and support from Bando per il Finanziamento della Ricerca Fondamentale 2022 dell'Istituto Nazionale di Astrofisica (INAF): GO Large program and from the Framework per l'Attrazione e il Rafforzamento delle Eccellenze (FARE) per la ricerca in Italia (R20L5S39T9). 

Computations were performed on the HPC system Raven at the Max Planck Computing and Data Facility. We acknowledge the project support by the Max Planck Computing and Data Facility. 
\end{acknowledgements}

\section*{Data Availability}\label{data_availability}

The lightcone generation code, \texttt{LightGen}, will be publicly available on GitLab\footnote{\url{https://github.com/SoumyaShreeram/LightGen/}} upon acceptance of this paper. The data underlying this article is available upon reasonable request to the corresponding author.

\bibliographystyle{aa} 
\bibliography{biblio}

\begin{appendix}

\end{appendix}

\end{document}